\pdfoutput=1
\documentclass[fleqn,usenatbib]{mnras}

\usepackage{newtxtext,newtxmath}

\usepackage[T1]{fontenc}

\DeclareRobustCommand{\VAN}[3]{#2}
\let\VANthebibliography\thebibliography
\def\thebibliography{\DeclareRobustCommand{\VAN}[3]{##3}\VANthebibliography}

\usepackage{graphicx}	
\usepackage{amsmath}	
\usepackage{siunitx}    
\usepackage{array}
\usepackage{multirow}
\usepackage{booktabs}
\usepackage{caption}

\newcommand{\borus}{\textsc{borus}}
\newcommand{\gro}{GRO J1655-40}
\newcommand{\gx}{GX 339-4}
\newcommand{\laedd}{$\lambda_{\mathrm{Edd}}$}

\newcommand{\mbh}{$M_\mathrm{BH}$}
\newcommand{\mcol}[1]{\multicolumn{1}{c}{#1}}

\newcommand{\msigma}{$M\text{--}\sigma_\star$}

\newcommand{\msun}{M$_{\sun}$}

\newcommand{\nbmc}{$N_\mathrm{BMC}$}

\newcommand{\nh}{$N_{\mathrm{H}}$}
\newcommand{\nustar}{{\it NuSTAR}}
\newcommand{\veldis}{$\sigma_\star$}
\newcommand{\xsvar}{$\sigma^2_{\mathrm{NXV}}$}

\newcommand{\xspec}{\textsc{xspec}}
\newcommand{\xte}{XTE J1550-564}

\title[Indirect methods for black hole masses]{Comparing indirect methods for black hole masses in AGN: the good, the bad, and the ugly}

\author[Gliozzi et al.]{
M. Gliozzi,$^{1}$\thanks{E-mail: mgliozzi@gmu.edu}
J. K. Williams,$^{1}$
A. Akylas,$^{2}$
I. E. Papadakis,$^{3}$
O. I. Shuvo,$^{4}$
A. Halavatkar,$^{1}$
A. Alt$^{1}$
\\
$^{1}$Department of Physics and Astronomy, George Mason University, 4400 University Drive, Fairfax, VA 22030, USA\\
$^{2}$Physics Department, University of Crete, 73010, Heraklion, Greece\\
$^{3}$Institute for Astronomy Astrophysics Space Applications and Remote Sensing (IAASARS), National Observatory of Athens,
I. Metaxa \& V. Pavlou, Penteli 15236, Greece\\
$^{4}$Department of Physics, University of Maryland Baltimore County, 1000 Hilltop Circle, Baltimore, MD 21250, USA
}

\date{Accepted XXX. Received YYY; in original form ZZZ}

\pubyear{2015}

\begin{document}
\label{firstpage}
\pagerange{\pageref{firstpage}--\pageref{lastpage}}
\maketitle

\begin{abstract}
The black hole mass \mbh\ is crucial in constraining the growth of supermassive BHs within their host galaxies. Since direct measurements of \mbh\ with dynamical methods are restricted to a limited number of nearly quiescent nearby galaxies and a small minority of active galactic nuclei (AGN), we must rely on indirect methods. In this work, we utilize an unbiased, volume-limited, hard X-ray selected sample of AGN to compare the reliability of some commonly used indirect methods, emphasising those that can be applied to obscured AGN.
Based on a subsample of AGN with \mbh\ determined via dynamical methods, our study suggests that  X-ray based techniques, such as the scaling method and the one based on the variability measured through the excess variance, are in good agreement with the dynamical methods.  On the other hand, the \msigma\ correlation based on inactive galaxies tends to systematically overestimate \mbh, regardless of the level of obscuration. We provide a correcting factor that produces an acceptable agreement with dynamical values and can be used to quickly correct the \mbh\ computed with this method. We also derive an alternative \msigma\ correlation based on this unbiased sample of AGN with a slope considerably shallower than the ones obtained using inactive galaxies, suggesting that the latter may not be appropriate to compute the \mbh\ in AGN. Finally, we find that no quick fix can be applied to correct the \mbh\ obtained from the fundamental plane of black hole activity, casting doubts on the reliability of this method. 
\end{abstract}




\section{Introduction}
It is now widely accepted that Active Galactic Nuclei (AGN), the most luminous persistent sources in the universe, are powered by accretion onto a supermassive black hole at the center of their host galaxy. Depending on the black hole mass, \mbh, and the accretion rate, $\dot m$, their bolometric luminosity $L_\mathrm{bol}$ may range from a tiny fraction of their host galaxy luminosity, as in the case of our Galaxy, to orders of magnitude beyond the host galaxy, as in the case of the most powerful quasars. Since the accretion rate is generally measured in terms of Eddington ratio, $\lambda_\mathrm{Edd}=L_\mathrm{bol}/L_\mathrm{Edd}$ (where the Eddington luminosity is defined as $L_\mathrm{Edd}=1.3\times 10^{38} (M_\mathrm{BH}/\mathrm{M_\odot})$ erg s$^{-1}$), the \mbh\ plays a crucial role in characterizing the intrinsic properties of AGN.

As for all astrophysical sources, the most reliable way to determine the mass of black holes is through dynamical methods: by measuring the orbital parameters of objects or gas gravitationally bound to the BH, one can constrain \mbh\ via the virial theorem (e.g., \citealt{Ferrarese2005}). For a few low-power AGN, close enough to have their gravitational sphere of influence spatially resolved by current instrumentation, the \mbh\ may be determined from stellar or gas dynamics measurements (e.g., \citealt{Gebhardt2003, Macchetto1997}). For brighter unobscured AGN (the so-called type 1 AGN), \mbh\ can be reliably constrained using the reverberation mapping (RM) technique, where the delay between the optical continuum, associated with the accretion disk, and the Balmer lines, produced in the broad line region (BLR), yields the size of the BLR, and the width of the Balmer lines constrains the BLR velocity (e.g., \citealt{Blandford1982, Peterson2004}). Unfortunately, the RM method is heavily resource- and time-consuming and can only be employed for broad-lined AGN that vary on a reasonably short amount of time, which severely limits its applicability. Luckily, the discovery of a tight correlation between the optical luminosity and the size of the BLR (e.g., \citealt{Kaspi2005, Bentz2006a}) has made it possible to extend the \mbh\ determination to a much larger number of broad-lined AGN, using the so-called single-epoch (SE) technique, where the measurement of the line-continuum delay is no longer necessary (e.g., \citealt{Laor1998, Wandel1999}). Although this indirect method has proven very successful in extending the estimate of the \mbh\ to thousands of broad-lined AGN, some caution should be considered when this technique is applied well beyond the range probed by the RM sample, such as at high redshift  (e.g., \citealt{Denney2012, Denney2016}) or in sources where there is substantial obscuration (see, e.g., \citealt{Mejia-Restrepo2022} and references therein).

Indeed, there is a major problem in accurately constraining \mbh\ in obscured AGN (type 2 AGN). In this case, the only direct dynamical method available is based on the rare detection of water masers in Keplerian motion within the accretion disk surrounding the BH. This technique has been successfully applied only to a fairly limited number of obscured AGN (e.g., \citealt{Greenhill2003, Kuo2011}). In addition, for a few partially obscured AGN with broad lines observable in the near-infrared band, a variation of the SE method, based on the correlation between BLR size and hard X-ray luminosity, was successfully utilized by \citet{Ricci2017b, Ricci2017c} and  \citet{Onori2017}. However, for the vast majority of obscured AGN, which represent the bulk of the AGN population, no simple direct method to constrain \mbh\ exists, and generally one has to rely on statistical correlations between \mbh\ and some properties associated with the bulge of the host galaxy, such as mass, luminosity, and velocity dispersion  $\sigma_\star$ (e.g., \citealt{Kormendy1995, Magorrian1998, Gebhardt2000, Ferrarese2000}). Since these correlations have been established using samples of mostly quiescent galaxies and unobscured AGN in the local universe, it is important to test their validity to a larger population of AGN with particular focus on the obscured ones.

In our recent work, using NGC 4151, one of the few AGN with \mbh\ determined with multiple direct dynamical methods,  we carried out a comparison between indirect methods that can be used to constrain \mbh\ in obscured AGN \citep{Williams2023}. Here, we expand our previous investigation by performing a systematic comparison between indirect methods using an unbiased sample that can be considered as representative of the AGN population in the local universe. More specifically, we will compare the \msigma\ correlation method \citep{Kormendy2013}, the fundamental plane of BH activity \citep{Gueltekin2019}, the X-ray scaling method \citep{Shaposhnikov2009, Gliozzi2011}, and the method based on X-ray variability as measured by the normalized excess variance (e.g., \citealt{Papadakis2004, Ponti2012, Akylas2022}). For completeness, we will also verify whether the optically based single epoch method is fully consistent with dynamical measurements for broad-lined AGN contained in our sample.
To this aim, we utilize a volume-limited sample of AGN, whose selection criteria -- very hard X-ray detection, short distance $D < 40$ Mpc, and X-ray bright with $L_\mathrm{X} > 10^{42}~{\rm erg~s^{-1}}$ -- ensure that only bona fide AGN that are unbiased in terms of obscuration are included, and that a sizable fraction of AGN has \mbh\ estimated with direct dynamical methods. Importantly, the near totality of the sample has been observed with \nustar\, which, thanks to its high sensitivity in the hard energy band (3--79 keV) and low background, is ideal to robustly constrain the spectral contributions of absorption and reflection and hence the primary emission, and to measure the intrinsic variability of the AGN.

The structure of the paper is the following. In Sections 2 and 3, we describe respectively the sample selection and the data reduction and analysis of \nustar\ data, as well as that of radio data. In Section 4, we report the \mbh\ values obtained with the different indirect methods, and in Section 5 we carry out a statistical analysis and systematic comparison of these methods. In Section 6, we discuss the main findings and draw our conclusions in Section 7. 

In brief, we found that the X-ray scaling method and the X-ray variability one yield \mbh\ values that are fully consistent with the dynamical ones. On the other hand the \mbh\ values obtained using the \msigma\ correlation of \citet{Kormendy2013} appear to be systematically overestimated. We provide a correction factor and derive an alternative \msigma\ correlation with a shallower slope that yields values consistent with the dynamical estimates of this volume-limited sample of AGN. We also confirm that the \mbh\ estimates derived with the FP are unreliable and there is not a simple correction that can make them consistent with the dynamical ones. Finally, for the broad-lined AGN of our sample, we verified that the optically based SE method provides values broadly consistent with the dynamical ones with a tendency to underestimate them in some (but not all) of the AGN that are substantially obscured.

\section{Sample selection}
\label{sec:sample} 
For our work we utilized AGN with hard X-ray luminosity $L_\mathrm{14-195 ~keV} > 10^{42}~{\rm erg~s^{-1}}$, taken from \citet{Oh2018}, and redshift-independent distance $D < 40$ Mpc from \citet{Koss2022a}.
Our sample was selected from the Swift 70-month Burst Alert Telescope (BAT) catalog \citep{Baumgartner2013}, where the detection of AGN at very hard X-ray energies (14--195 keV) ensures that the sample is almost independent of obscuration. Our choice of a hard X-ray luminosity threshold guarantees that all the selected objects are genuine AGN, and the distance criterion makes sure that the sample is not biased towards the brightest AGN. When dealing with nearby AGN, one must consider the potential error associated with the distance, since their actual values can be substantially different from the distances obtained from their radial velocity assuming a uniform expansion and Hubble's law. Luckily, the second data release of the Swift BAT AGN Spectroscopic Survey (BASS DR2, \citealt{Koss2022a}) provides high-quality redshift-independent distances for all nearby AGN, determined following the approach of \citet{Leroy2019}, which we adopted as bona fide distances.

Following these criteria our sample is made up of 15 broad-lined AGN with five of them classified as Seyfert 1.9 galaxies (hereafter, Sy1.9), whose optical spectra are characterized by a narrow H$_\beta$ and a broad H$_\alpha$ lines, and 20 narrow-lined AGN (hereafter, Sy2). After excluding the sources without archival \nustar\ data (NGC 4235 and LEDA 89913) and one source that is likely to be jet-dominated (Cen A), our resulting sample is made up of 32 objects. 

Since our goal is to test the reliability of \mbh\ estimates obtained with various indirect methods, we searched the literature for direct dynamical measurements and found 11 objects whose \mbh\ was determined either via maser measurements (five objects), gas dynamics (two objects), or reverberation mapping with direct modeling of the BLR via velocity-delay maps (four objects). These 11 objects define our {\it restricted dynamical sample}, which comprises the most reliable  \mbh\ estimates to be compared with the values obtained with different indirect methods. Ideally, we would compare all the indirect methods tested in this work using this restricted sample. However, some of these methods (namely, the X-ray variability method and the single epoch method based on broad H$_{\alpha}$ line measurements) have \mbh\ estimates for less than ten objects in the restricted dynamical sample. For this reason, we also defined an {\it extended dynamical sample} that comprises six additional AGN (for a total of 17 objects with dynamical mass), of which four have the \mbh\ determined with the single epoch method in the near-IR \citep{Ricci2017b} and two objects with the traditional reverberation mapping, where a virial factor ${\langle f\rangle}=4.3$ was adopted. 
In the remainder of the paper, we will explicitly state when a result refers to the restricted dynamical sample or to the extended dynamical sample.

The basic characteristics (name, BAT id, classification, distance, dynamically determined \mbh, and relative reference) of our AGN sample are reported in Table \ref{tab:sample}.
\begin{table*}
	\caption{Sample properties}		
	\begin{center}
	\begin{tabular}{rllrrll} 
			\toprule
			\toprule       
			\mcol{BAT ID} & \mcol{Name} &  \mcol{Type} & \mcol{Distance} & \mcol{\mbh} & \mcol{Method} &
            \mcol{Reference} \\
			  & & & \mcol{(Mpc)} & \mcol{(\msun)} & & \\
			\midrule
			93 & NGC 678 & Sy2 & 34.50 & \dots & \dots & \dots \\
			140 & NGC 1052$^{**}$ & Sy2 & 19.23 & 
            $4.27 \times 10^6$ & IR virial &
            \citet{Onori2017} \\
			144 & NGC $1068^*$ & Sy1.9 & 14.40 & 
            $8.51 \times 10^6$ & maser &
            \citet{Lodato2003} \\
			184 & NGC $1365^*$ & Sy1 & 19.57 & 
            $4.38 \times 10^6$ & gas dynamics &
            \citet{Combes2019} \\
            216 & NGC 1566 & Sy1 & 17.90 & 
             \dots & \dots & \dots \\
			308 & NGC 2110 & Sy2 & 34.30 & \dots
             & \dots & \dots \\
			319 & ESO 5-4 & Sy2 & 28.18 & 
            \dots & \dots & \dots \\
			471 & NGC $2992^{**}$ & Sy1.9 & 38.00 & 
            $5.25 \times 10^6$ & IR virial &
            \citet{Onori2017} \\
            480	& NGC 3081 & Sy2 & 32.50
            & \dots & \dots & \dots \\
            484 & NGC $3079^*$ & Sy2 & 20.61 & 
            $2.86 \times 10^6$ & maser &
            \citet{Kondratko2005} \\
            497 & NGC $3227^{**}$ & Sy1 & 22.95 & 
            $4.84 \times 10^6$ & RM &
            \citet{DeRosa2018} \\
            530 & NGC $3516^{**}$ & Sy1 & 38.90 & 
            $2.50 \times 10^7$ & RM &
            \citet{DeRosa2018} \\
            558 & NGC $3783^*$ & Sy1 & 38.50 & 
            $2.82 \times 10^7$ & BLR model &
            \citet{Bentz2021} \\
            595 & NGC $4151^*$ & Sy1 & 15.80 & 
            $1.66 \times 10^7$ & BLR model &
            \citet{Bentz2022} \\
            615 & NGC 4388$^*$ & Sy2 & 18.11 & 
            $8.01 \times 10^6$ & maser &
            \citet{Kuo2011} \\
            621 & NGC 4500 & Sy2 & 34.51 & 
            \dots & \dots & \dots \\
            631 & NGC $4593^*$ & Sy1 & 37.20 & 
            $4.47 \times 10^6$ & BLR model &
            \citet{Williams2018} \\
            655 & NGC $4945^*$ & Sy2 & 3.72 & 
            $1.41 \times 10^6$ & maser &
           \citet{Greenhill1997} \\
            688 & NGC 5290 & Sy2 & 34.51 & 
            \dots & \dots & \dots \\
            711 & Circinus$^*$ & Sy2 & 4.21 & 
            $1.70 \times 10^6$ & maser &
            \citet{Greenhill2003} \\
            712 & NGC 5506 & Sy1.9 & 26.40 & 
            \dots & \dots & \dots \\
            739 & NGC 5728 & Sy1.9 & 37.50 & 
            \dots & \dots & \dots \\
            823 & ESO 137-34 & Sy2 & 34.10 & 
            \dots & \dots & \dots \\
            838 & NGC $6221^{**}$ & Sy2 & 11.86 & 
            $2.88 \times 10^6$ & IR virial &
            \citet{Onori2017} \\
            875 & NGC 6300 & Sy2 & 13.18 & 
            \dots & \dots & \dots \\
            1046 & NGC $6814^*$ & Sy1 & 22.80 & 
            $2.63 \times 10^6$ & BLR model &
            \citet{Pancoast2014} \\
            1135 & NGC 7172 & Sy2 & 33.90 & 
            \dots & \dots & \dots \\
            1142 & NGC 7213 & Sy1 & 22.00 & 
            \dots & \dots & \dots \\
            1157 & NGC $7314^{**}$ & Sy1.9 & 16.75 & 
            $1.74 \times 10^6$ & IR virial &
            \citet{Onori2017} \\
            1180 & NGC 7465 & Sy2 & 27.20 &\dots & \dots & \dots \\
            1184 & NGC 7479 & Sy2 & 36.81 & 
            \dots & \dots & \dots \\
            1188 & NGC 7582$^*$ & Sy2 & 22.49 & 
            $5.35 \times 10^7$ & gas dynamics &
            \citet{Wold2006} \\
			\bottomrule
		\end{tabular}	
	\end{center}
	\begin{flushleft}
		A single asterisk after the name denotes our restricted dynamical sample of 11 objects. A double asterisk denotes the sources added to the restricted sample to form our extended sample of 17 objects. The source for the distances is \citet{Koss2022a}. All \mbh\ values obtained with methods that depend on the distance were renormalized using the distances quoted in this table.
	\end{flushleft}
	\label{tab:sample}
\end{table*}

\section{Data reduction}
\label{sec:data}

We processed the entire sample with the \nustar\ data analysis script \textsc{nupipeline}, which calibrates and screens the data. Our version of the calibration database CALDB is 20200429. We defined each source region as a circle centered on the source and each background region as a circle nearby in the same image in an area that did not appear to contain source photons. Depending on the brightness of the source, our circles had radii from 30 to 100 arcsec. 
We used both focal plane modules FPMA and FPMB on the satellite to produce spectra and light curves with the \textsc{nuproducts} script. We list all the \nustar\ observations used in this paper in Table \ref{tab:obsids} in the Appendix, where we report the effective exposures (which by definition are shorter than the observation durations). In the spectral analysis for sources with multiple \nustar\ exposures, we chose the observations with longer exposure, higher mean count rate, and relatively low variability. The first two criteria favor higher signal-to-noise spectra, whereas the latter reduces the possibility of spectral variability affecting the time-average spectra (see \citealt{Williams2023} for more details). For the temporal analysis, following \citet{Akylas2022} all the observations with length longer than 80 ks were utilized.

\subsection{X-ray spectral analysis}
\label{sec:X-ray spectral}
One of the \mbh\ estimation methods we wish to test is the X-ray scaling method, which is based on the spectral fitting of the primary X-ray emission of the objects in our sample with the bulk Comptonization model \texttt{BMC} \citep{Titarchuk1997}.
To accurately constrain the primary X-ray emission produced in the corona in heavily obscured sources (such as the majority of our sample), it is important to properly parameterize the contributions of absorption and reflection associated with the torus. This goal is achieved by using the broadband X-ray spectrum afforded by \nustar\ and employing physically motivated models of the torus. As in our previous works focused on sources that are part of this sample \citep{Gliozzi2021,Shuvo2022}, we used as baseline model 
\begin{verbatim}
phabs * (atable(Borus) + MYTZ*BMC)
\end{verbatim}
where the first absorption model \texttt{phabs} parameterizes the Galaxy contribution, the \borus\ table model describes the continuum scattering and fluorescent emission line components associated with the torus \citep{Balokovic2018}, and the zeroth-order component of MYTorus, \texttt{MYTZ}, models the absorption and Compton scattering that act on the transmitted primary emission \citep{Murphy2009}. As explained in more detail in \citet{Gliozzi2021}, we use \texttt{MYTZ} because it models the Compton scattering with the Klein-Nishina cross section, which plays an important role  at energies above 10 keV in heavily absorbed AGN. 
The Comptonization model \texttt{BMC} has four parameters: the normalization \nbmc, the spectral index $\alpha$ (linked to the photon index by $\Gamma=\alpha+1$), the temperature of the seed photons $kT$, and $\log A$, which is related to the fraction of scattered seed photons $f$ by the relationship $A=f/(1 - f)$ (note that $\log A$ has very little impact on the determination of \mbh). 

To maintain the self-consistency of the torus model, the spectral index of the \textsc{bmc} model was linked to the photon index value of the torus model and \nbmc\ was forced to be equal to the \borus\ power-law normalization divided by a factor of 30, based on the empirical relationship obtained in \citet{Gliozzi2021}. 

For the few objects with relatively low absorption, the \texttt{MYTZ} component is substituted by a \texttt{zphabs} model left free to vary. For a few sources with more complex spectra, additional components were added to the baseline model; in particular, we included a model describing the fraction of primary emission that is directly scattered towards the observer by an optically thin ionized medium (\texttt{const*bmc}). 

The main spectral parameters obtained by fitting this baseline model and the 2--10 keV luminosity are reported in Table~\ref{tab:spectral} in the Appendix, along with the details on the sources that required more complex models, which were not already published in our previous works. We performed the X-ray spectral analysis using the \xspec\ \texttt{v.12.9.0} software package \citep{Arnaud1996}, and  the errors quoted on the spectral parameters represent the 1$\sigma$ confidence level. 

\subsection{X-ray temporal analysis}
\label{sec:X-ray temporal}
Another indirect method that we want to test is based on X-ray variability. For this reason, we extracted \nustar\ light curves in the 10--20 keV band and estimated the normalised excess variance based on the prescription presented in \citet{Akylas2022}. 
In brief, the light curves obtained for the FMPA and FMPB modules were first background subtracted and then summed using the {\sc LCMATH} tool within {\sc FTOOLS}.
To measure the variability power of the sources in our sample, we compute the  normalised excess variance \cite[e.g.,][]{nandra1997} using:

\begin{equation}
{\rm
\sigma_{NXV}^{2}=\frac{1}{N\mu^2} \sum_{i=1}^{N} \left( (X_i - \mu )^2 -\sigma_i^2 \right),}
\label{snvx}
\end{equation}

\noindent where $N$ is the number of bins in the light curve, $\rm X_i$ and $\rm \sigma_i$ are the count rates and their uncertainties, respectively, and $\mu$ is the unweighted  mean count rate.  

The $\sigma_{NXV}^{2}$ has been measured over light curve segments with a duration of $\rm \Delta t=$ 80 ks. When more than one valid segment is available for a source, we compute the mean of the individual excess variances.

\subsection{Radio analysis}
\label{sec:radio}
For the majority of our sources, we either utilized the 5 GHz VLA peak intensities (F$_\mathrm{peak}$) provided by \citet{Fisher2021} or used the C-band ($\sim4.89$~GHz) VLA archival\footnote{\url{http://www.aoc.nrao.edu/~vlbacald/src.shtml}} images to measure the peak intensities, whenever available in A-Configuration\footnote{\url{https://public.nrao.edu/vla-configurations/}}. Similarly to \citet{Williams2023}, we used VLA (kpc scale) measurements in our analysis because AGN do not follow the fundamental plane of black hole activity when radio intensities are measured on milli-arcsecond (subparsec) scales from VLBA observations \citep{Fisher2021, Shuvo2022}. The declination coverage of VLA is restricted to $\delta >$ $-$48$^{\circ}$; therefore, for some of the sources below this threshold (NGC 1566, ESO 5-4, NGC 6221, NGC 6300) we have used images at 4.8 GHz from the 64-meter Parkes radio telescope, and for a few others (NGC 4945, NGC 7213, Circinus) we have instead used images at 5 GHz from the Australia Telescope Compact Array (ATCA), which are  available on the NASA Extragalactic Database (NED)\footnote{\url{https://ned.ipac.caltech.edu}}. We could not find any archival images at 5 GHz for NGC 678, NGC 5290, or ESO 137$-$34.

We used the NRAO Common Astronomical Software Applications, also known as \textsc{casa} \citep[][]{McMullin2007}\footnote{\url{https://casa.nrao.edu}}, to analyze the images obtained from the VLA archive or NED. In \textsc{casa}, we utilized the two-dimensional fitting application through the ``viewer'' command to determine the peak intensities directly from the image. The radio luminosities (erg s$^{-1}$) shown in column 7 of Table~\ref{tab:mbhresults} for all sources were calculated using the redshift-independent distances provided in Table~\ref{tab:sample}.

\section{Black hole mass measurements}
\label{sec:MBH_measurements}
In the vast majority of AGN, dynamical methods cannot be applied either because they are too distant and hence their BH sphere of influence is too small to be spatially resolved or because the central engine and more specifically the BLR is not directly observable. In those cases, indirect methods have a crucial importance. Here, we briefly describe a few of these indirect methods whose accuracy and reliability will be tested against dynamical mass estimates in the following section. We will mainly focus on methods that can be applied to obscured AGN, that is, objects for which the BLR is not detected or hardly detected. However, for comparison purposes, we also test the optically based single epoch method. The black hole mass values obtained with the different indirect methods are reported in Table \ref{tab:mbhresults}.

\subsection{Single epoch spectra}
\label{sec:SE}
The single epoch (SE) method is the most utilized method to estimate \mbh\ of broad-lined AGN at all redshifts. The \mbh\ is obtained from the same formula used in the reverberation mapping method:

\begin{equation}\label{eq_vir}
M_{\mathrm{BH}}= f \frac{\Delta V^2 R}{G} 
\end{equation}
where $\Delta V$ is the velocity dispersion of a specific broad emission line (generally measured from full width at half maximum or from the standard deviation $\sigma$ of the Gaussian model used to fit the line), $R$ is the size of the BLR, $G$ the universal gravitational constant, and $f$ the virial factor, which encompasses the unknown geometry and inclination of the BLR.

The key difference between SE and RM is that in the latter case $R$ is obtained from the time delay between variations of the optical/UV ionizing continuum and those observed in the BLR ($R=c \Delta t$), whereas in the SE method the BLR size is directly obtained from the tight correlation between $R$ and a monochromatic continuum luminosity in the optical band $L_{\rm opt}$, which was empirically derived from reverberation mapping studies (e.g., \citealt{Kaspi2005, Bentz2006a}). 

This method has been successfully applied to thousands of broad-lined AGN. However, some caution should be used since the average virial factor $\langle f \rangle$ introduces a systematic uncertainty of $\sim$ 0.4 dex (e.g., \citealt{Pancoast2014}). Additionally, although the correlation $R\textrm{--}L_{\rm opt}$ was established for a nearby sample of RM AGN, the same correlation is extended to very distant AGN using high-ionization lines like C\,\textsc{iv}, which are often blue-shifted because of winds and hence not reliable tracers of the BLR dynamics (e.g., \citealt{Trakhtenbrot2012}). Finally, it has been recently suggested that the optically based SE method tends to underestimate the \mbh\ in AGN with substantial obscuration (e.g., \citealt{Caglar2020, Mejia-Restrepo2022}). For these reasons, we have included the SE method in our statistical comparison with the dynamical methods.

To be more specific, in this work we compute the \mbh\ via the SE method using the broad H\,${\upalpha}$ values of FWHM(H\,$\upalpha$) and L(H\,$\upalpha$) provided by \citet{Mejia-Restrepo2022} in their table 5, as well as their equation 2:
\begin{equation}\label{eq_SE}
M_{\mathrm{BH}}= 2.67\times10^6 \left( \frac{L(\mathrm{H}\,\upalpha)}{10^{42} ~ \mathrm{erg ~ s^{-1}}} \right)^{0.55} \left( \frac{\mathrm{FWHM}(\mathrm{H}\,\upalpha)}{10^3 ~ {\mathrm{km ~ s^{-1}}}} \right)^{2.06} \mathrm{M}_\odot
\end{equation}

\subsection{M--\texorpdfstring{\veldis}{sigma} relation}
\label{sec:Msigma}
Over the last few decades, numerous studies focusing on nearby galaxies have revealed the existence of several relatively tight correlations between properties of the galaxy bulge (for example, luminosity, mass, and velocity dispersion \veldis) and the mass of the supermassive BH at the center of the galaxy, suggesting a co-evolution between these two elements operating at very different scales (see, e.g., \citealt{Kormendy2013}, for a detailed review). Since \msigma\ is the tightest among these correlations, it has been frequently used to estimate \mbh\ in the absence of more direct dynamical methods.

In our recent work, where we compared different indirect methods using  NGC 4151 as a laboratory, we tested different versions of this correlation \citep{Williams2023}. More specifically, we compared the dynamical value of \mbh\ with those derived with the \msigma\ correlation of \citet{McConnell2013}, which makes a distinction between early and late type galaxies, with the equation of \citet{Woo2013}, which was calibrated using an RM sample of AGN, with the correlation of \citet{Kormendy2013}, which was derived from a sample of galaxies with classical bulges and with \mbh\ determined with direct dynamical methods, and finally with the correlation proposed by \citet{Shankar2016}, who, assuming that previous correlations were biased towards higher values of \veldis, proposed a new equation that they named the intrinsic correlation. Of all the \msigma\ correlations tested, only the one proposed by \citet{Kormendy2013} yielded a value in agreement with the dynamical estimate of NGC 4151. For this reason, and because
the same correlation is used by \citet{Koss2022b} to estimate the \mbh\ for the vast majority of the AGN in BASS DR2,
throughout this work we will use the \citet{Kormendy2013} correlation, which is described by the equation

\begin{equation}
	\label{eq:McConnell}
    \log \left( \frac{M_{\mathrm{BH}}}{\mathrm{M}_\odot} \right) = 8.49 + 4.38 \log \left( \frac{\sigma_\star}{200\, \mathrm{km\, s}^{-1}} \right)
\end{equation}

\subsection{Fundamental plane of BH activity}
\label{sec:FP}
The fundamental plane of black hole activity (hereafter, FP) is a correlation involving X-ray luminosity, radio luminosity, and \mbh, which was independently proposed by \citet{Merloni2003} and \citet{falcke2004} to unify the accretion and ejection properties of black hole systems at all scales. When other methods to constrain \mbh\ are unavailable, the FP has been frequently used in AGN studies, despite its inherent large scatter (nearly 1 dex), which yields order of magnitude estimates rather than tight constraints on \mbh. 

As for the \msigma\ correlation, different versions of the FP were developed over the years and three of them were tested in \citet{Williams2023}, where the FP proposed by \citet{Dong2014} for relatively highly accreting BH systems yielded the closest estimate to the dynamical mass of NGC 4151. However, when we tried to apply this FP to our volume-limited sample of AGN, we obtained unreasonable values with \mbh\ ranging from a few solar masses to $10^{11}\ {\rm M}_\odot$. We therefore used the FP recently proposed by \citet{Gueltekin2019}, which was derived using only dynamically constrained \mbh\ and is described by the equation below.

\begin{equation}
	\label{eq:Gueltekin}
    \mu=(0.55\pm0.22) + (1.09\pm0.10)R+\left(-0.59^{+0.16}_{-0.15}\right)X   
\end{equation}
where $\mu=\log(M_\mathrm{BH}/10^8\; \mathrm{M}_\odot)$, $R=\log(L_\mathrm{5~GHz}/10^{38}~\mathrm{erg\,s}^{-1})$, and $X=\log( L_\mathrm{2-10~keV}/10^{40}~\mathrm{erg\,s}^{-1})$.

Although high-resolution VLBA data exist for most of the AGN in our sample, we utilized lower-resolution VLA data for our estimate of \mbh\ based on the FP, because it was recently shown that  using higher-resolution radio data the FP breaks down \citep{Fisher2021, Shuvo2022}.

\subsection{X-ray scaling method}
\label{sec:Xscaling}
The X-ray scaling method, introduced by \citet{Shaposhnikov2009} for X-ray binary systems (XRBs) with an accreting BH, was first successfully applied  to unobscured AGN with \mbh\ determined via RM \citep{Gliozzi2011} and then extended to heavily obscured AGN obtaining \mbh\ consistent with those determined via maser measurements \citep{Gliozzi2021}, making it one of the few truly universal methods, since it can be equally utilized for stellar mass BHs and supermassive ones, as well as for obscured and unobscured BH systems. 

In brief, this method determines \mbh\ in AGN by scaling up the dynamically measured mass of a stellar-mass BH (the reference source) under the assumption that the primary X-ray emission from the corona is produced by the same Comptonization process in AGN and XRBs and that $\Gamma$ can be considered a faithful proxy of the accretion rate of all BH systems.  A detailed description of the method is provided in \citet{Gliozzi2021} and references therein; here we just report the equation used to determine the \mbh. A derivation of this equation is provided in \citet{Williams2023}, where the tabulated values of \nbmc\ for the different reference sources are also provided.

\begin{equation}\label{eq_Xscal1}
M_{\mathrm{BH,AGN}}=M_{\mathrm{BH,ref}}  \times \left(\frac{N_{\mathrm{BMC,AGN}}}{N_{\mathrm{BMC,ref}}}\right)  \times \left(\frac{d_{\mathrm{AGN}}^2}{d_{\mathrm{ref}}^2}\right)
\end{equation}
where \nbmc\ is the normalization of the Comptonization model \textsc{bmc} model (measured in units of $10^{39}$ erg s$^{-1}$ divided by the distance squared in units of 10 kpc) and $d$ the distance in kpc. 

The application of this method to NGC 4151 in \citet{Williams2023} confirms the conclusions derived in previous works that there are two reference sources, \gro\ during the 2005 decay and \gx\ during the 2003 decay, which yield the most reliable estimates of \mbh. For this reason, throughout this work we computed the \mbh\ with both reference sources (when available) and then calculated the average, which is the value reported in the table. Only one source (Circinus) has $\Gamma$ too steep  for these reference sources. In that case, we utilized \xte\ during the 1998 rise as reference source, and multiplied the mass estimate by a factor of 4, which is the correction needed for this reference source to be consistent with dynamically measured \mbh\ \citep{Williams2023}.

\begin{figure} 
 \includegraphics[width=\columnwidth]{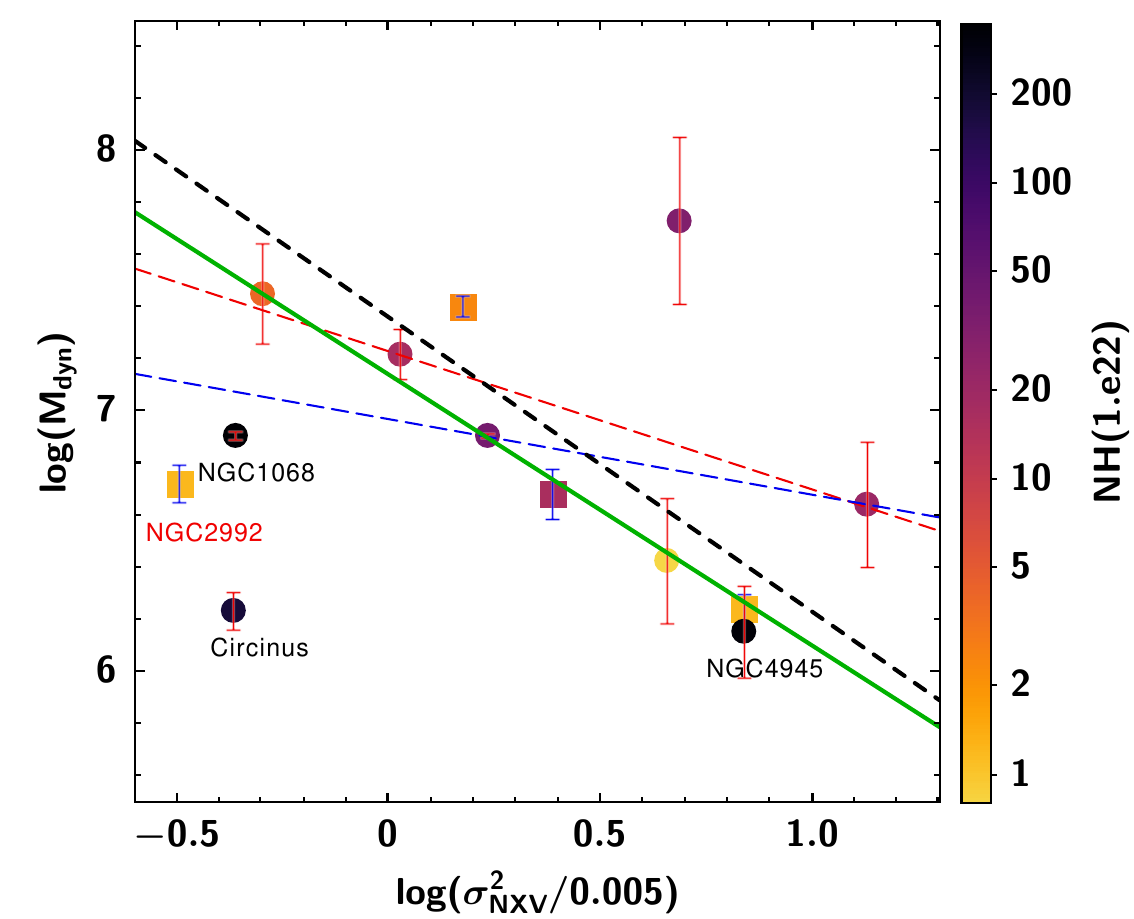}
\caption{\mbh\ plotted vs. \xsvar\ in the 10--20 keV band. The symbols are color coded based on the intrinsic \nh\ along the line of sight; circles indicate objects of the dynamical restricted sample, whereas squares are the added objects in the extended sample. Compton-thick (CT) sources (NGC 1068, NGC 4945, Circinus) have been excluded from the fit but are shown in the plot. The green continuous line represents the best fit, $\log M_{\rm BH}=-1.04 \times \log(\sigma^2_{\rm NXV}/0.005)+7.14$, obtained from the extended dynamical sample without the changing-look AGN NGC 2992, whose low variability may be affected by its peculiar state and hence unreliable. The short-dashed black line represents the best fit, $\log M_{\rm BH}=-1.13 \times \log(\sigma^2_{\rm NXV}/0.005)+7.38$, from \citet{Akylas2022} using reverberation mapping data only. For completeness, we also show
the best fit, $\log M_{\rm BH}=-0.53 \times \log(\sigma^2_{\rm NXV}/0.005)+7.23$, obtained using only the restricted dynamical sample (long-dashed red line), which is limited to six objects (after excluding CT AGN and objects without reliable \xsvar), and the best fit, $\log M_{\rm BH}=-0.29 \times \log(\sigma^2_{\rm NXV}/0.005)+6.97$, obtained using the extended dynamical sample (ten objects; blue long-dashed line).}
  \label{fig:logMdlogNXV}
\end{figure}
\subsection{X-ray variability method}
\label{sec:Xvas}
\begin{table*}
	\caption{\mbh\ results}		
	\begin{center}
	\begin{tabular}{lrrrrrrrrrr} 
			\toprule
			\toprule
			\mcol{Name} & \mcol{FWHM} & 
            \mcol{log $L$} & \mcol{\mbh} & 
            \mcol{$\sigma_\star$} & \mcol{\mbh} & 
            \mcol{$L_\text{5~GHz}$} & \mcol{\mbh} & 
            \mcol{$\sigma^2_\mathrm{NXV}$} & \mcol{\mbh} & 
            \mcol{\mbh} \\
			  & \mcol{bH$\upalpha$} & \mcol{bH$\upalpha$} & 
            \mcol{SE} & & 
            \mcol{\msigma} & & 
            \mcol{FP} & & \mcol{Xvar} & 
            \mcol{Xscal} \\
			  & \mcol{(km s$^{-1}$)} & & 
            \mcol{(\msun)} & \mcol{(km s$^{-1}$)} & 
            \mcol{(\msun)} & \mcol{(erg s$^{-1}$)} & 
            \mcol{(\msun)} & & \mcol{(\msun)} & 
            \mcol{(\msun)} \\
			  \mcol{(1)} & \mcol{(2)} & \mcol{(3)} & \mcol{(4)} & 
            \mcol{(5)} & \mcol{(6)}& \mcol{(7)} & \mcol{(8)} & 
            \mcol{(9)} & \mcol{(10)} & \mcol{(11)} \\
			\midrule
			NGC 678 & \dots & \dots & 
            \dots & 179 &
            $1.90\!\times\!10^8$ & \dots & 
            \dots & \dots & 
            \dots & $6.91_{-2.49}^{+7.23}\!\times\!10^5$ \\
            \noalign{\smallskip}
			NGC 1052 & \dots & \dots & 
            \dots & 237 &
            $6.50\!\times\!10^8$ & $3.9\!\times\!10^{39}$ &
            $2.00\!\times\!10^9$ & \dots & \dots & 
            $1.03_{-0.34}^{+0.76}\!\times\!10^6$ \\
            \noalign{\smallskip}
            NGC 1068 & 2165 & 41.60 & 
            $7.90\!\times\!10^6$ & 158 &
            $1.10\!\times\!10^8$ & $3.0\!\times\!10^{38}$ & 
            $3.9\!\times\!10^7$ & $2.19\!\times\!10^{-3}$ & 
            $3.27\!\times\!10^7$ & $8.00_{-1.03}^{+0.84}\!\times\!10^6$ \\
            \noalign{\smallskip}
            NGC 1365 & 1095 & 39.37 & 
            $1.15\!\times\!10^5$ & 141 &
            $6.68\!\times\!10^7$ & $1.4\!\times\!10^{38}$ & 
            $5.0\!\times\!10^7$ & $6.74\!\times\!10^{-2}$ & 
            $9.22\!\times\!10^5$ & $1.68_{-0.53}^{+1.14}\!\times\!10^6$ \\
            \noalign{\smallskip}
            NGC 1566 & 1914 & 40.74 & 
            $2.06\!\times\!10^6$ & 97.7 &
            $1.34\!\times\!10^7$ & $1.5\!\times\!10^{38}$ & 
            $4.3\!\times\!10^7$ & $1.69\!\times\!10^{-2}$ & 
            $3.90\!\times\!10^6$ & $8.08_{-3.06}^{+4.86}\!\times\!10^5$ \\
            \noalign{\smallskip}
            NGC 2110 & \dots & \dots & 
            \dots & 233 &
            $6.03\!\times\!10^8$ & $3.7\!\times\!10^{38}$ & 
            $1.14\!\times\!10^7$ & \dots & 
            \dots & $1.04_{-0.31}^{+0.61}\!\times\!10^8$ \\
            \noalign{\smallskip}
            ESO 5-4 & \dots & \dots & 
            \dots & 113 &
            $2.53\!\times\!10^7$ & $1.4\!\times\!10^{38}$ & 
            $8.78\!\times\!10^7$ & \dots & 
            \dots & $2.40_{-1.28}^{+0.47}\!\times\!10^6$ \\
            \noalign{\smallskip}
            NGC 2992 & 2255 & 40.56 & 
            $2.30\!\times\!10^6$ & 152 &
            $9.29\!\times\!10^7$ & $6.7\!\times\!10^{37}$ & 
            $3.1\!\times\!10^6$ & $1.60\!\times\!10^{-3}$ & 
            $4.51\!\times\!10^7$ & $2.12_{-0.61}^{+0.98}\!\times\!10^7$ \\
            \noalign{\smallskip}
            NGC 3081 & \dots & \dots & 
            \dots & 130 &
            $4.68\!\times\!10^7$ & $3.1\!\times\!10^{36}$ & 
            $1.52\!\times\!10^5$ & $1.22\!\times\!10^{-2}$ & 
            $5.47\!\times\!10^6$ & $1.50_{-0.41}^{+0.72}\!\times\!10^7$ \\
            \noalign{\smallskip}
            NGC 3079 & \dots & \dots & 
            \dots & 154 &
            $9.84\!\times\!10^7$ & $1.4\!\times\!10^{38}$ & 
            $9.99\!\times\!10^6$ & \dots & 
            \dots & $8.03_{-2.18}^{+2.57}\!\times\!10^6$ \\
            \noalign{\smallskip}
            NGC 3227 & 2859 & 40.56 & 
            $3.75\!\times\!10^6$ & 126.8 &
            $4.20\!\times\!10^7$ & $3.3\!\times\!10^{37}$ & 
            $4.7\!\times\!10^6$ & $1.22\!\times\!10^{-2}$ & 
            $5.45\!\times\!10^6$ & $4.14_{-1.16}^{+1.72}\!\times\!10^6$ \\
            \noalign{\smallskip}
            NGC 3516 & 4133 & 40.94 & 
            $1.30\!\times\!10^7$ & 153.6 &
            $9.72\!\times\!10^7$ & $2.3\!\times\!10^{37}$ & 
            $1.4\!\times\!10^6$ & $7.50\!\times\!10^{-3}$ & 
            $9.06\!\times\!10^6$ & $1.18_{-0.81}^{+0.46}\!\times\!10^7$ \\
            \noalign{\smallskip}
            NGC 3783 & 2824 & 41.59 & 
            $1.35\!\times\!10^7$ & 113 &
            $2.53\!\times\!10^7$ & $7.2\!\times\!10^{37}$ & 
            $5.0\!\times\!10^6$ & $2.54\!\times\!10^{-3}$ & 
            $2.80\!\times\!10^7$ & $1.53_{-0.48}^{+0.56}\!\times\!10^7$ \\
            \noalign{\smallskip}
            NGC 4151 & 2855 & 41.51 & 
            $1.25\!\times\!10^7$ & 91.8 &
            $1.02\!\times\!10^7$ & $5.7\!\times\!10^{37}$ & 
            $3.6\!\times\!10^6$ & $5.32\!\times\!10^{-3}$ & 
            $1.29\!\times\!10^7$ & $1.30_{-0.42}^{+0.60}\!\times\!10^7$ \\
            \noalign{\smallskip}
            NGC 4388 & \dots & \dots & 
            \dots & 99 &
            $1.42\!\times\!10^7$ & $6.0\!\times\!10^{36}$ & 
            $1.58\!\times\!10^6$ & $8.53\!\times\!10^{-3}$ & 
            $7.92\!\times\!10^6$ & $4.74_{-1.45}^{+3.26}\!\times\!10^6$ \\
            \noalign{\smallskip}
            NGC 4500 & \dots & \dots & 
            \dots & 134 &
            $5.35\!\times\!10^7$ & $3.5\!\times\!10^{36}$ & 
            $4.40\!\times\!10^7$ & \dots & 
            \dots & $2.16_{-1.47}^{+1.11}\!\times\!10^6$ \\
            \noalign{\smallskip}
            NGC 4593 & 3539 & 40.57 & 
            $5.90\!\times\!10^6$ & 122.3 &
            $3.58\!\times\!10^7$ & $9.3\!\times\!10^{36}$ & 
            $7.9\!\times\!10^5$ & \dots & 
            \dots & $5.97_{-1.62}^{+1.92}\!\times\!10^6$ \\
            \noalign{\smallskip}
            NGC 4945 & \dots & \dots & 
            \dots & 114 &
            $2.63\!\times\!10^7$ & $1.3\!\times\!10^{38}$ & 
            $1.24\!\times\!10^6$ & $3.47\!\times\!10^{-2}$ & 
            $1.84\!\times\!10^6$ & $7.57_{-2.01}^{+3.24}\!\times\!10^5$ \\
            \noalign{\smallskip}
            NGC 5290 & \dots & \dots & 
            \dots & 141 &
            $6.68\!\times\!10^7$ & \dots & 
            \dots & \dots &
            \dots & $2.33_{-0.85}^{+0.12}\!\times\!10^6$ \\
            \noalign{\smallskip}
            Circinus & \dots & \dots & 
            \dots & 101 &
            $1.55\!\times\!10^7$ & $3.2\!\times\!10^{37}$ & 
            $1.09\!\times\!10^7$ & $2.14\!\times\!10^{-3}$ &
            $3.33\!\times\!10^7$ & $2.06_{-0.52}^{+0.84}\!\times\!10^6$ \\
            \noalign{\smallskip}
            NGC 5506 & 1758 & 40.22 & 
            $8.96\!\times\!10^5$ & 104 &
            $1.76\!\times\!10^7$ & $4.1\!\times\!10^{38}$ & 
            $3.90\!\times\!10^7$ & $5.14\!\times\!10^{-3}$ &
            $1.34\!\times\!10^7$ & $1.13_{-0.33}^{+0.42}\!\times\!10^7$ \\
            \noalign{\smallskip}
            NGC 5728 & 1820 & 40.52 & 
            $1.41\!\times\!10^6$ & 176 &
            $1.77\!\times\!10^8$ & $3.0\!\times\!10^{37}$ & 
            $2.60\!\times\!10^6$ & \dots &
            \dots & $9.93_{-3.41}^{+7.03}\!\times\!10^6$ \\
            \noalign{\smallskip}
            ESO 137-34 & \dots & \dots & 
            \dots & 118 &
            $3.06\!\times\!10^7$ & \dots & 
            \dots & \dots &
            \dots & $7.13_{-3.12}^{+18.2}\!\times\!10^6$ \\
            \noalign{\smallskip}
            NGC 6221 & \dots & \dots & 
            \dots & 65 &
            $2.25\!\times\!10^6$ & $1.3\!\times\!10^{38}$ & 
            $1.17\!\times\!10^8$ & \dots &
            \dots & $2.10_{-0.77}^{+1.18}\!\times\!10^5$ \\
            \noalign{\smallskip}
            NGC 6300 & \dots & \dots & 
            \dots & 81 &
            $5.90\!\times\!10^6$ & $3.9\!\times\!10^{37}$ & 
            $9.21\!\times\!10^6$ & \dots &
            \dots & $1.30_{-0.47}^{+0.52}\!\times\!10^6$ \\
            \noalign{\smallskip}
            NGC 6814 & 3286 & 40.9 & 
            $7.69\!\times\!10^6$ & 108.1 &
            $2.09\!\times\!10^7$ & $3.5\!\times\!10^{36}$ & 
            $3.90\!\times\!10^5$ & $2.28\!\times\!10^{-2}$ &
            $2.86\!\times\!10^6$ & $3.18_{-0.84}^{+1.02}\!\times\!10^6$ \\
            \noalign{\smallskip}
            NGC 7172 & \dots & \dots & 
            \dots & 167 &
            $1.40\!\times\!10^8$ & $8.0\!\times\!10^{36}$ & 
            $3.53\!\times\!10^5$ & \dots &
            \dots & $1.40_{-0.41}^{+0.81}\!\times\!10^7$ \\
            \noalign{\smallskip}
            NGC 7213 & 2312 & 40.61 & 
            $2.58\!\times\!10^6$ & 158.8 &
            $1.13\!\times\!10^8$ & $3.9\!\times\!10^{38}$ & 
            $1.10\!\times\!10^8$ & $1.80\!\times\!10^{-2}$ &
            $3.63\!\times\!10^6$ & $1.37_{-0.54}^{+0.55}\!\times\!10^6$ \\
            \noalign{\smallskip}
            NGC 7314 & 1264 & 39.42 & 
            $1.65\!\times\!10^5$ & 63 &
            $1.96\!\times\!10^6$ & $2.8\!\times\!10^{36}$ & 
            $4.10\!\times\!10^5$ & $3.48\!\times\!10^{-2}$ &
            $1.84\!\times\!10^6$ & $1.68_{-0.23}^{+0.17}\!\times\!10^6$ \\
            \noalign{\smallskip}
            NGC 7465 & \dots & \dots & 
            \dots & 85 &
            $7.28\!\times\!10^6$ & $5.2\!\times\!10^{36}$ & 
            $8.70\!\times\!10^5$ & \dots &
            \dots & $1.85_{-0.50}^{+0.59}\!\times\!10^6$ \\
            \noalign{\smallskip}
            NGC 7479 & \dots & \dots & 
            \dots & 109 &
            $2.16\!\times\!10^7$ & $1.9\!\times\!10^{37}$ & 
            $4.69\!\times\!10^6$ & \dots &
            \dots & $5.07_{-2.37}^{+12.3}\!\times\!10^6$ \\
            \noalign{\smallskip}
            NGC 7582 & \dots & \dots & 
            \dots & 95 &
            $1.19\!\times\!10^7$ & $8.3\!\times\!10^{36}$ & 
            $9.30\!\times\!10^5$ & $2.44\!\times\!10^{-2}$ &
            $2.65\!\times\!10^6$ & $5.03_{-1.43}^{+2.25}\!\times\!10^6$ \\
			\bottomrule
		\end{tabular}	
	\end{center}
	\begin{flushleft}
		Columns. 1 = object name. 2 = full width at half maximum for the broad H\,$\upalpha$ line. 3 = log of the luminosity of the broad H\,$\mathrm{\upalpha}$ line. 4 = \mbh\ computed with the single epoch method. 5 = stellar velocity dispersion in km s$^{-1}$. 6 = \mbh\ computed with the \msigma\ method of \citet{Kormendy2013}. 7 = radio luminosity at 5 GHz in erg s$^{-1}$. 8 = \mbh\ computed with the FP method of \citet{Gueltekin2019}. 9 = normalized excess variance computed with 80~ks segments in the 10--20~keV band. 10 = \mbh\ computed with the X-ray variability method. 11 = \mbh\ computed with the X-ray scaling method.
	\end{flushleft}
	\label{tab:mbhresults}
\end{table*}

High persistent variability is one of the defining characteristics of AGN at all wavelengths. However, on relatively short timescales, it is particularly prominent in the X-ray band, which has the additional advantages with respect to the optical/UV bands of being less affected by absorption and more easily separated from star/galaxy contributions. The normalised excess variance \xsvar, estimator of the intrinsic
`band-variance' of the source, \citep[e.g.,][]{nandra1997} has been extensively used  as a predictor of the \mbh\ of AGN. Past studies have demonstrated the existence of a tight anti-correlation between the normalized excess variance \xsvar\ and \mbh\ in both soft (0.5--2 keV) and hard (2--10 keV) energy ranges (e.g., \citealt{Papadakis2004,Ponti2012,Lanzuisi2014}). Very recently, \citet{Akylas2022} extended this technique to the harder X-ray band afforded by \nustar, making easier the application of this technique to obscured AGN. To this end, \citet{Akylas2022} utilized a combined sample of AGN with \mbh\ determined via RM and the \msigma\ correlation method and explored in detail the minimum requirements for the light curves that will be used to compute \xsvar, so that the resulting BH mass estimates will be as unbiased as possible and of known variance. 

Since in this work we want to assess the reliability of several indirect methods (including the \msigma\ correlation), and since the anti-correlation between \mbh\ and \xsvar\ derived by \citet{Akylas2022} is based on black hole mass values derived from the \msigma\ correlation and from the reverberation mapping technique (RM), we will only use the anti-correlation obtained with the RM sample. However, since the RM sample itself is calibrated on the \msigma\ correlation, we also derive a new anti-correlation using the \mbh\ values of our extended dynamical sample. The use of the extended sample is necessary because the three Compton-thick sources (NGC 1068, NGC 4945, Circinus) and the changing-look AGN NGC 2992 with an anomalously low value of \xsvar\ are excluded from the correlation analysis, further reducing the number of data points. Our results are shown in Fig.~\ref{fig:logMdlogNXV}, where the green continuous line represents our best fit, which is similar but slightly offset with respect to the best fit obtained by \citet{Akylas2022} using reverberation mapping data only (short-dashed black line). For completeness, in this figure we also show the best fit obtained using the restricted sample of six objects (long-dashed red line) and that from the extended dynamical sample including the CL AGN (blue long-dashed line).

In the remainder of the paper, when we discuss the \mbh\ estimated with the X-ray variability technique, we will quote both the ones obtained using the RM sample in \citet{Akylas2022} (whose anti-correlation is described by the short-dashed black line in Fig.~\ref{fig:logMdlogNXV}) and those derived from the anti-correlation obtained with the extended dynamical sample in this work (whose anti-correlation is represented by the continuous green line in the same figure).

\section{Comparison of indirect methods}
\label{sec:comparison} 

\begin{figure*}
\includegraphics[width=5.5cm]{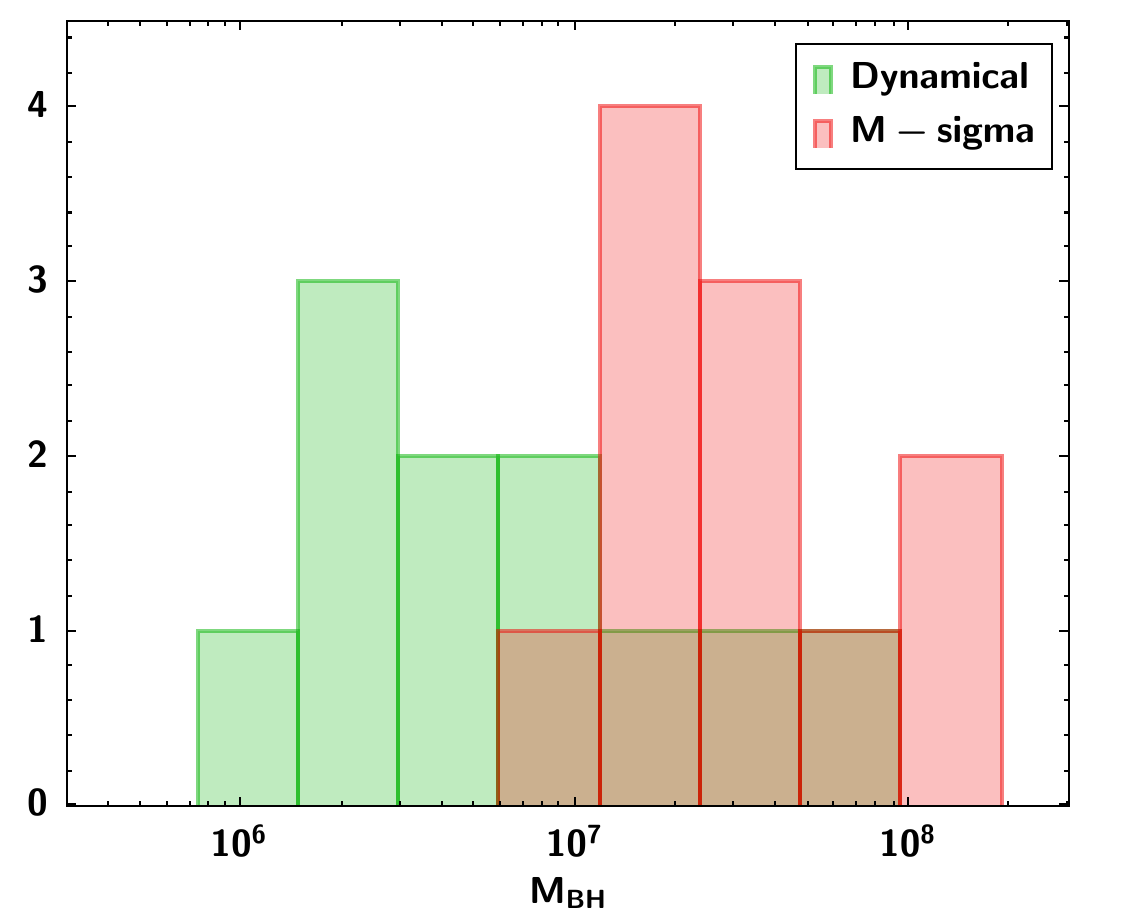}\includegraphics[width=5.5cm]{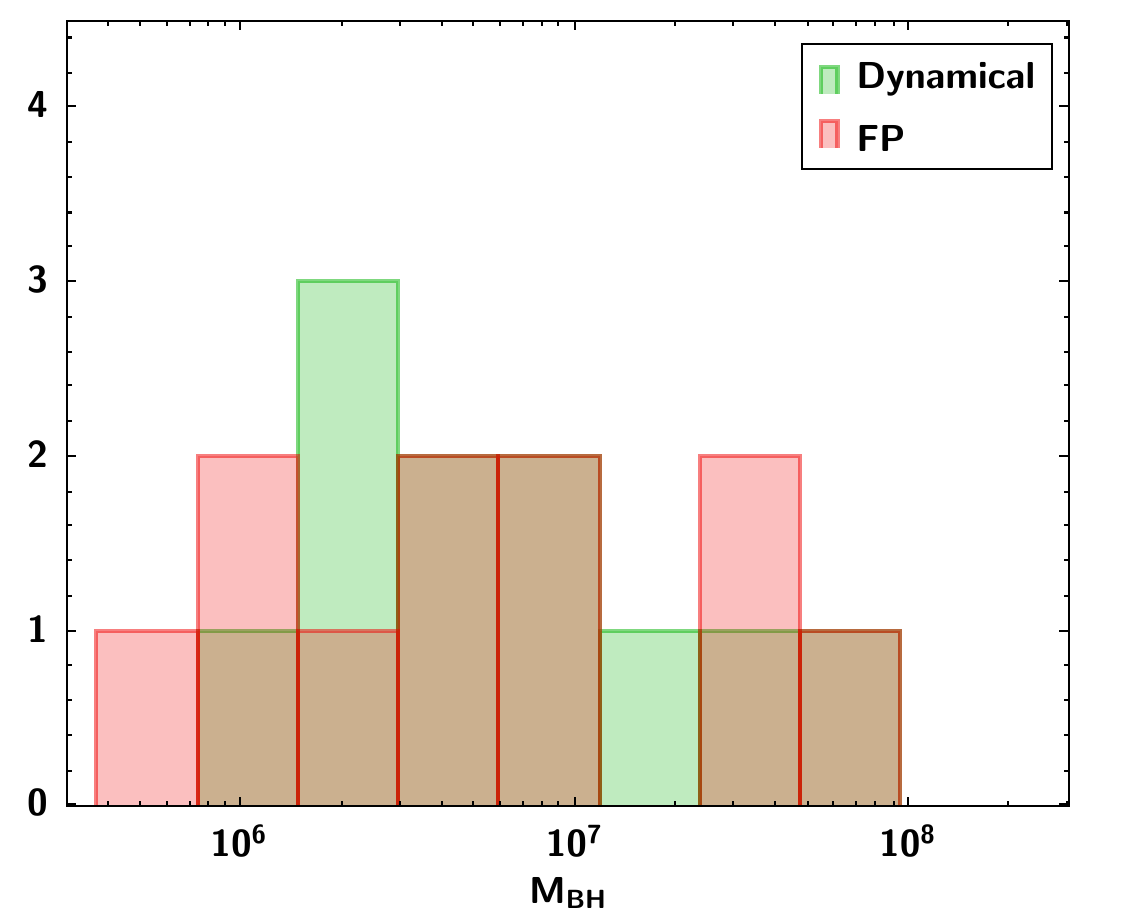}\includegraphics[width=5.5cm]{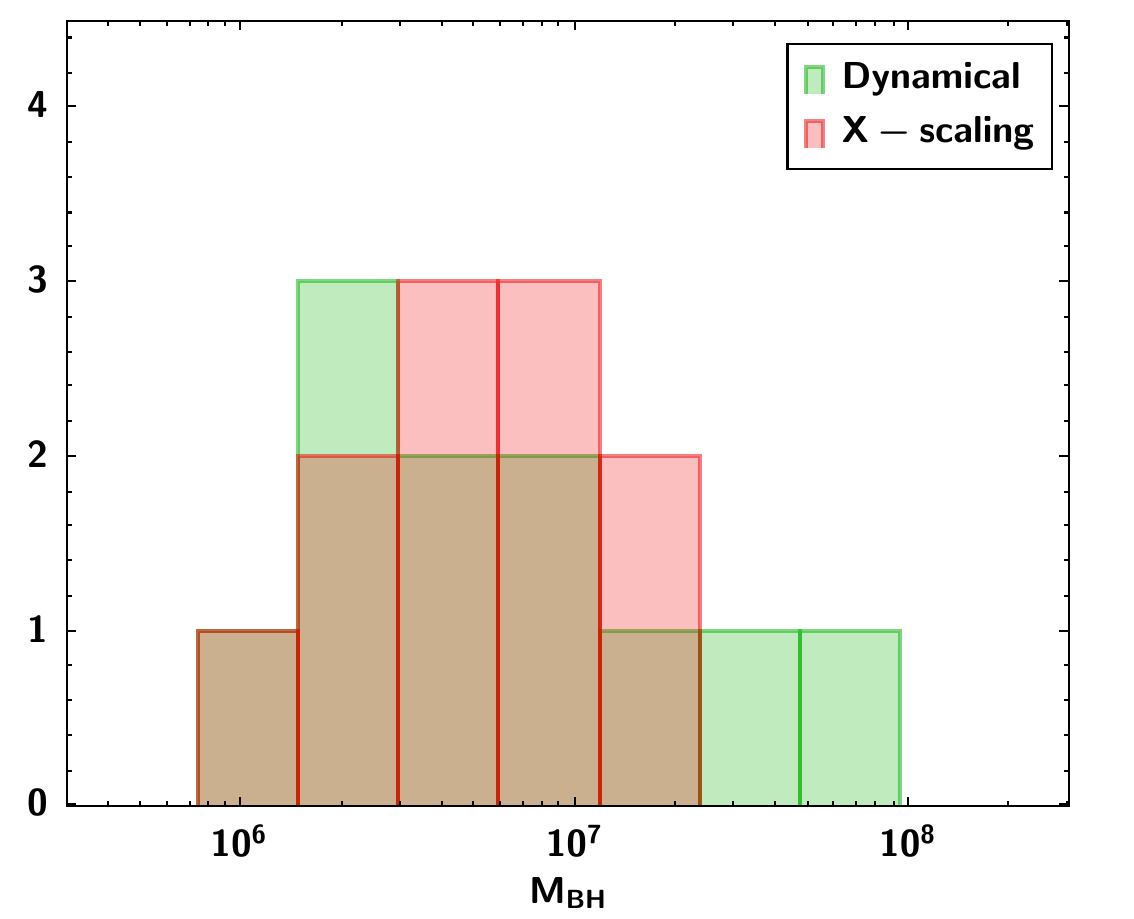}
\includegraphics[width=5.5cm]{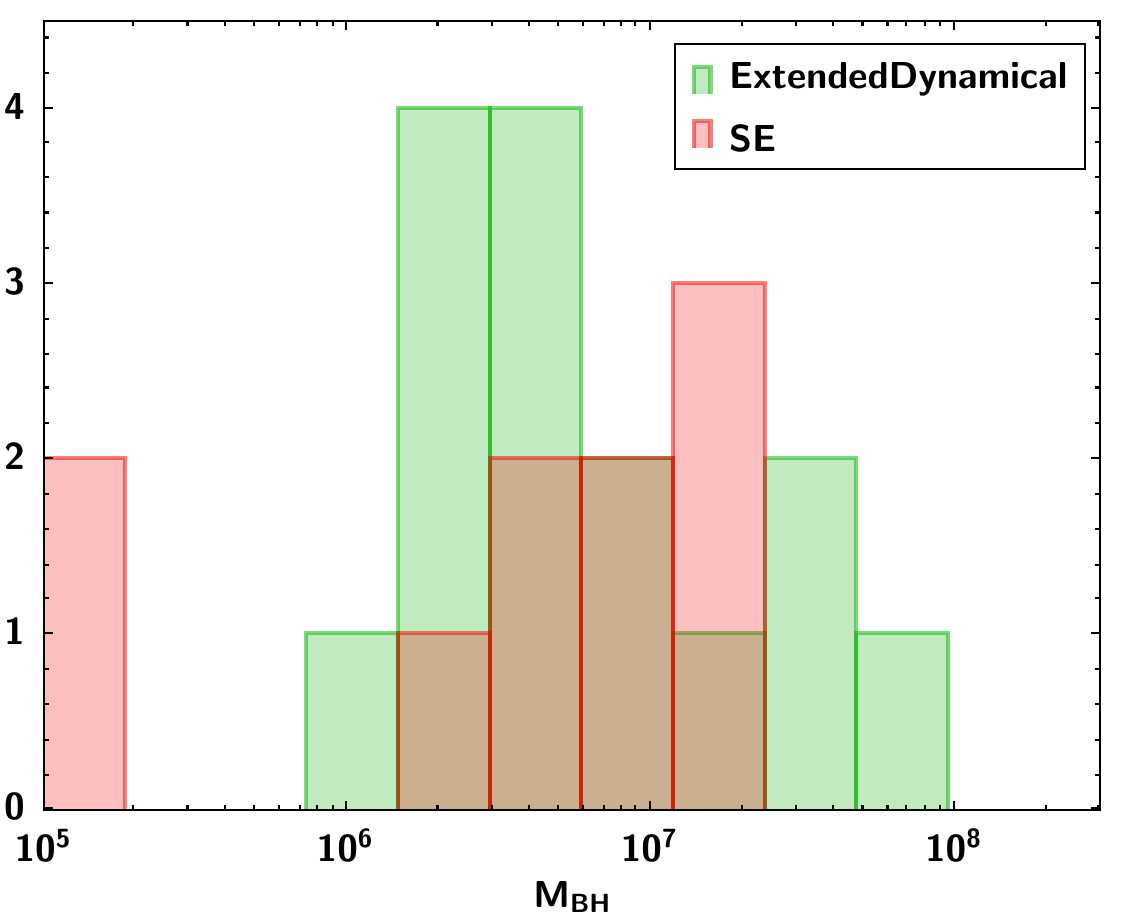}\includegraphics[width=5.5cm]{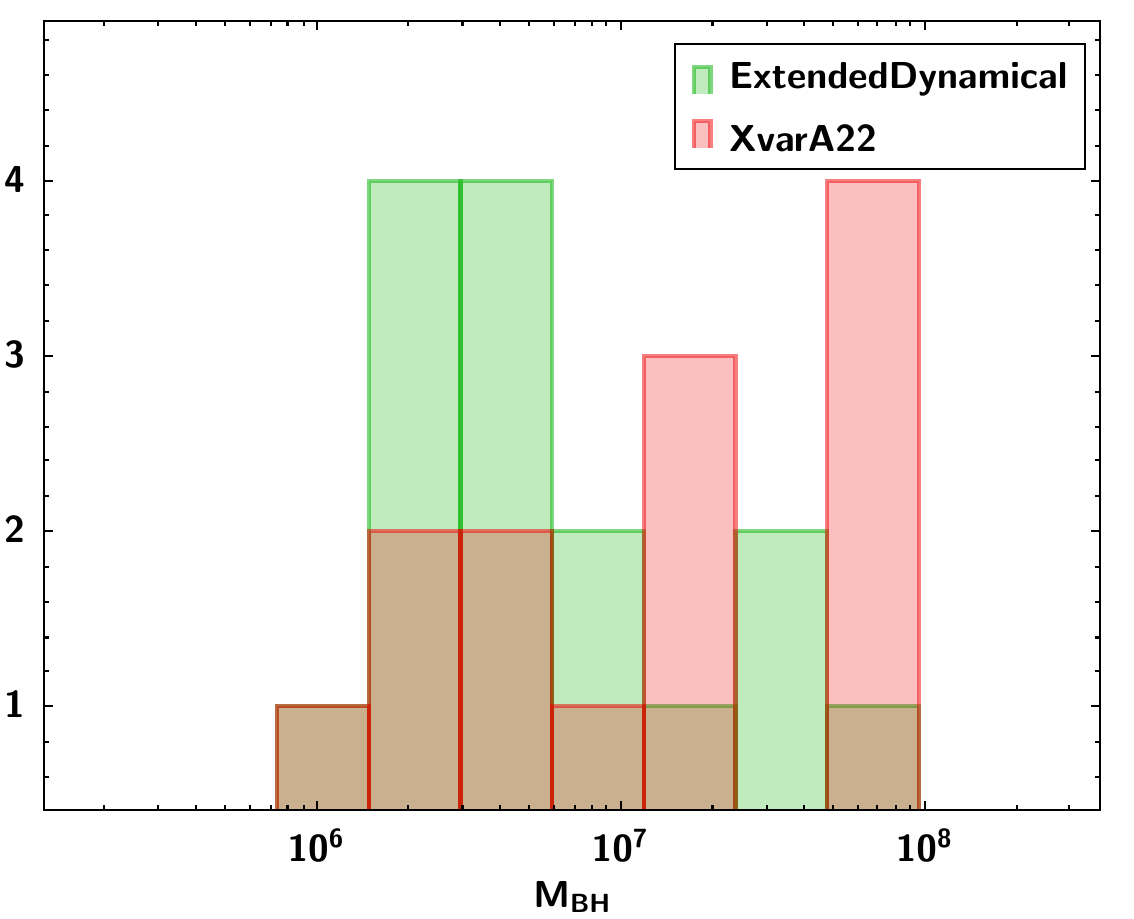}\includegraphics[width=5.5cm]{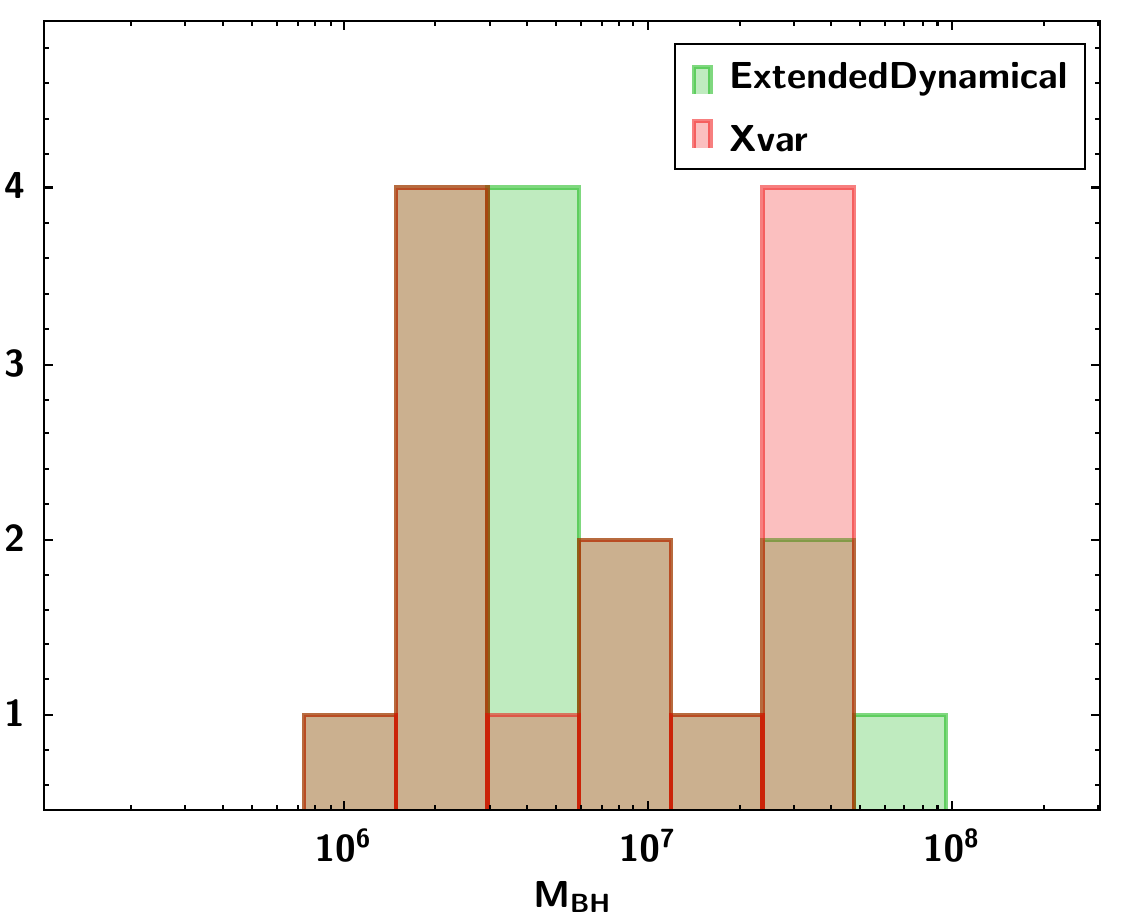}
\caption{Top panels: comparison between the restricted dynamical \mbh\ distribution and those obtained with the \msigma\ correlation (left), with the fundamental plane of black hole activity (middle), and with the X-ray scaling method (right). Bottom panels: comparison between the extended dynamical \mbh\ distribution and the ones obtained with 
single epoch H\,$\upalpha$ measurements (left), with the X-ray variability method based on the correlation derived by \citet{Akylas2022} using the RM sample (middle), and the variability method based on the correlation derived in this work (right).   
}
\label{fig:histog}
\end{figure*}

\begin{table*}
	\centering
	\caption{Statistical comparisons of indirect methods}
	\label{tab:indirect_methods}
	\begin{tabular}{ccccccc}
		\hline
        \hline
		Method & Dynamical sample & $P_{\rm K-S}$ &  $P_{\rm t}$ & $r$ ($P_{\rm S}$) & $\tau$ ($P_{\rm K}$)  & $\langle {\rm max}(M_{\rm BH}\, {\rm ratio}) \rangle $ ($\sigma/\sqrt{n}$)\\
		\hline
		\noalign{\smallskip}
   \msigma  & R (11) & $2.5 \times 10^{-3}$ & $3 \times 10^{-2}$ &  $-0.37$ (0.25)             & $-0.24$(0.31)                & 10.6 (3.0)\\
    FP          & R (11) & 0.74                 & 0.68               & $-0.35$ (0.28)              & $-0.24$ (0.31)               & 13.0 (5.0)\\
   X-scal      & R (11) & 0.99                 & 0.26               & 0.69 ($2 \times 10^{-2}$)& 0.56 ($2 \times 10^{-2}$)      & 2.5 (0.8)\\
   \hline
   SE          & E (10) & 0.97                 & 0.28               & 0.80 ($5 \times 10^{-3}$) &  0.64 ($9 \times 10^{-3}$) & 6.3 (3.6)\\
    X-var-80ks  & E (13) & 0.99                 & 0.77               & 0.28 (0.36)               & 0.19 (0.35)               & 5.2 (1.9)\\
    X-var$_{\rm A22}$-80ks       & E (13) & 0.49                 & 0.17               & 0.27 (0.37)               & 0.18 (0.39)               & 6.8 (2.7)\\
		\hline
	\end{tabular}
	\begin{flushleft}
		Columns 1 = indirect method, 2 = dynamical sample (R means restricted and E extended) used for the comparison with the number of sources available in parentheses, 3 = probability of the Kolmogorov--Smirnov test (large values indicate that the \mbh\ distribution obtained with the indirect method is indistinguishable from that obtained from direct dynamical methods), 4 = probability of the Student's $t$ test (small values indicate that the mean of the distributions compared are significantly different), 5 = Spearman's rank correlation coefficient $r$ and relative probability, 6 = Kendall's rank correlation coefficient $\tau$ and relative probability, 7 = average of the maximum between $M_{\rm BH,dyn}/M_{\rm BH,indirect}$ and $M_{\rm BH,indirect}/M_{\rm BH,dyn}$ with the standard deviation divided by the square root of the number of objects in parentheses.
  
	\end{flushleft}
\end{table*}

\begin{figure*}
\includegraphics[width=5.5cm]{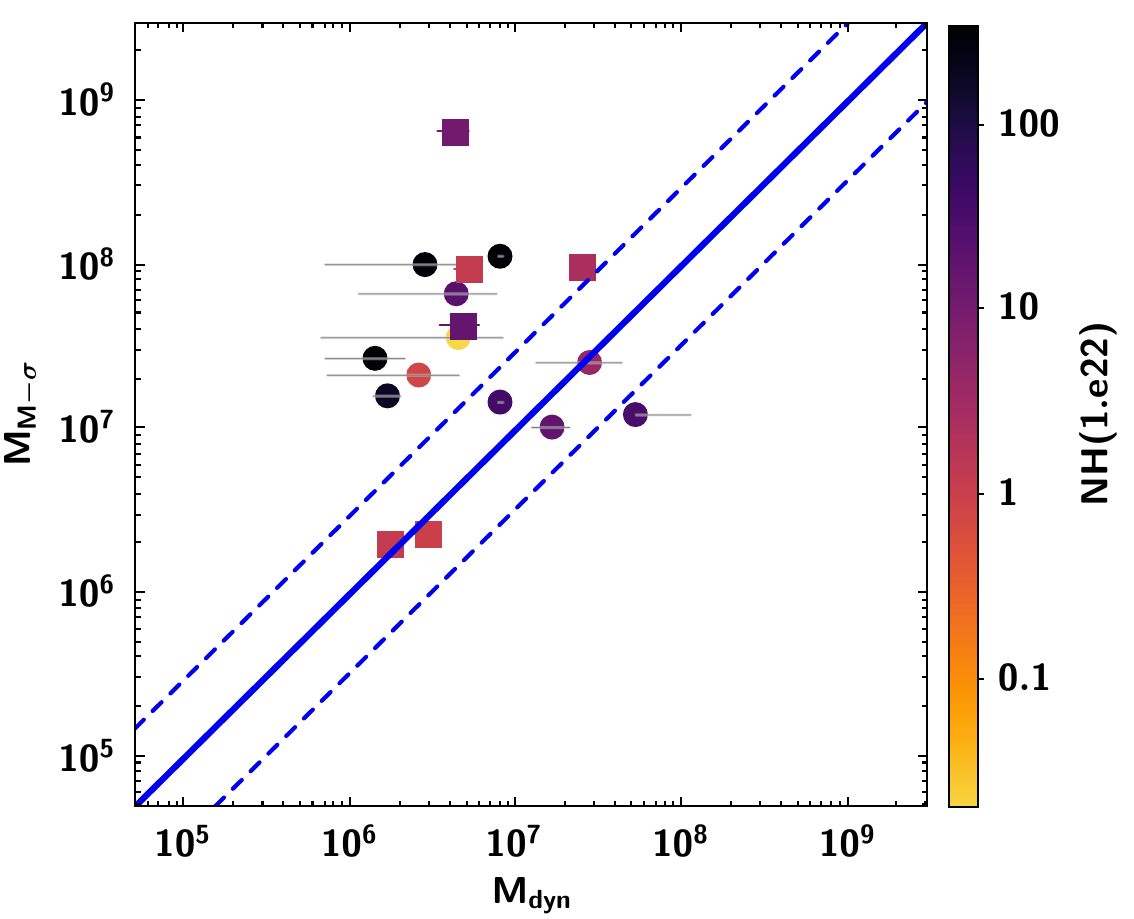}\includegraphics[width=5.5cm]{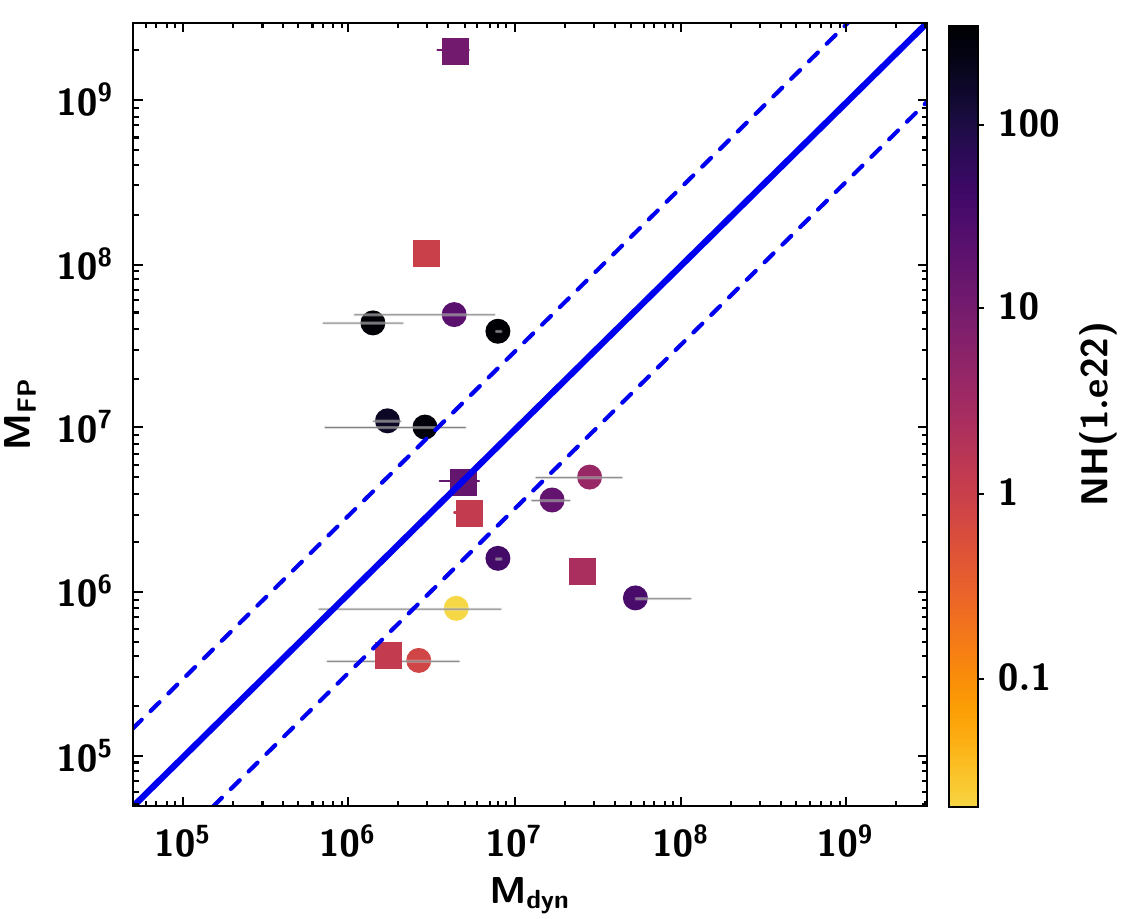}\includegraphics[width=5.5cm]{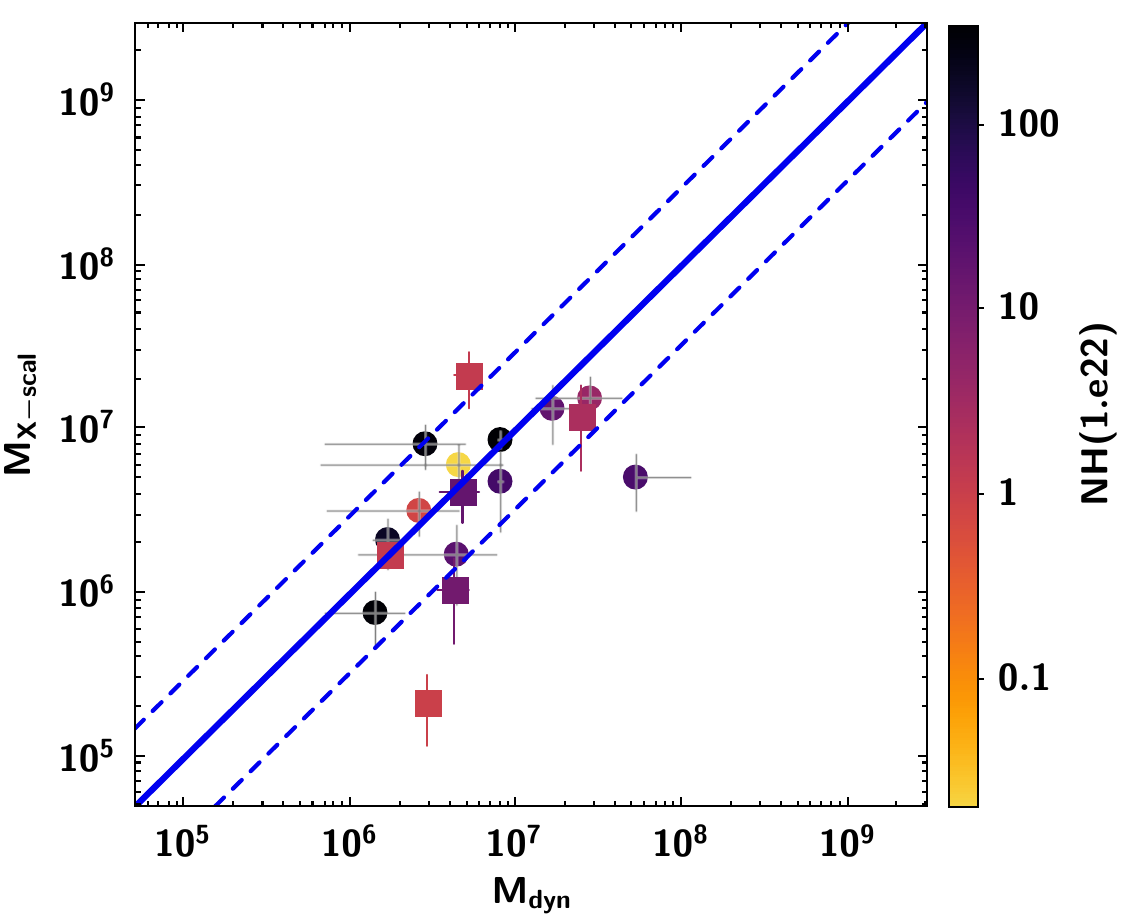}
\includegraphics[width=5.5cm]{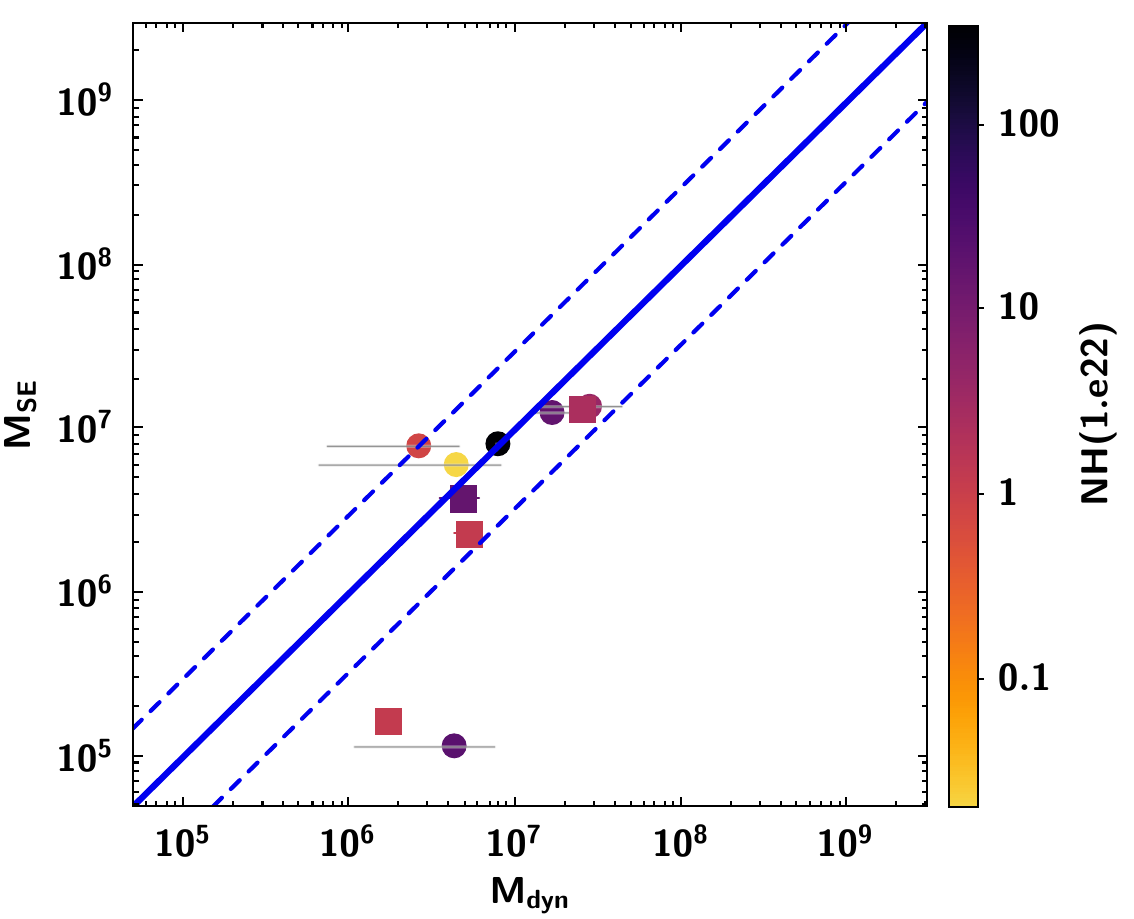}\includegraphics[width=5.5cm]{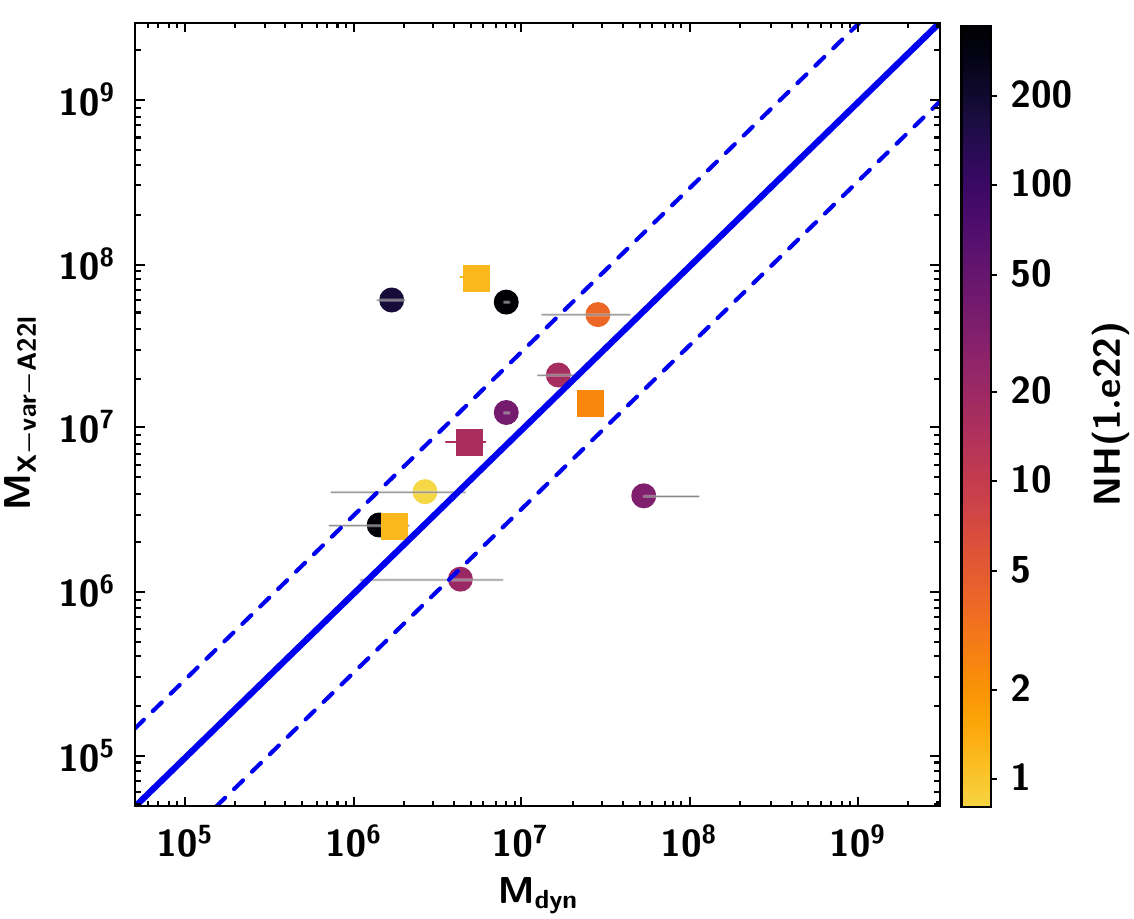}\includegraphics[width=5.5cm]{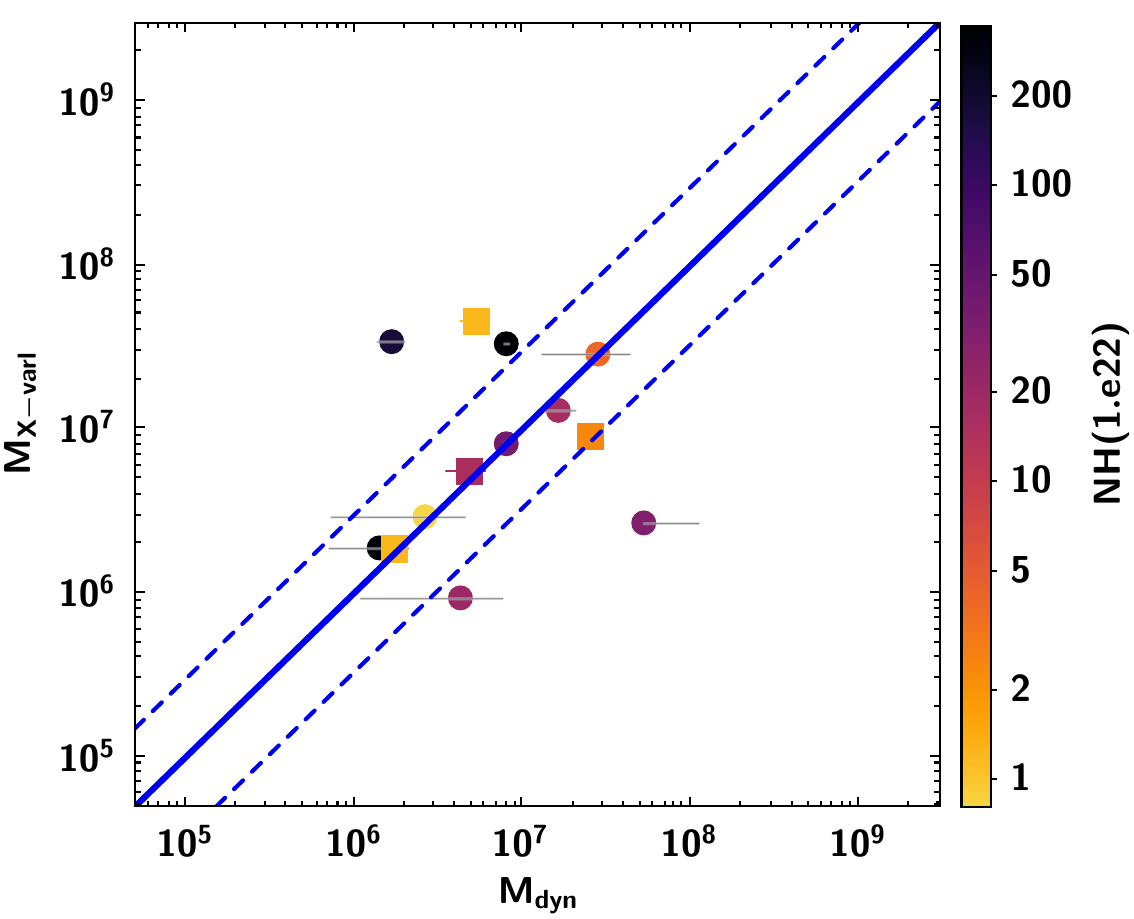}
\caption{Top left: \mbh\ obtained from the \msigma\ correlation plotted vs. those obtained from dynamical methods (circles indicates objects from the restricted sample, whereas squares represent the sources added to form the extended dynamical sample). Top middle: \mbh\ from the fundamental plane of black hole activity vs. dynamical values. Top right:  values from the X-ray scaling method vs. the dynamical ones. Bottom left: \mbh\ obtained with single epoch H\,$\upalpha$ measurements plotted vs. those obtained from dynamical methods. Bottom middle and right: \mbh\ obtained from X-ray variability using the correlation of \citet{Akylas2022} and that obtained in this work, respectively, plotted vs. the dynamical measurements. The symbols are color coded based on the intrinsic \nh\ along the line of sight. 
}
\label{fig:xy_dyn_indir}
\end{figure*}

\begin{figure}
\includegraphics[width=\columnwidth]{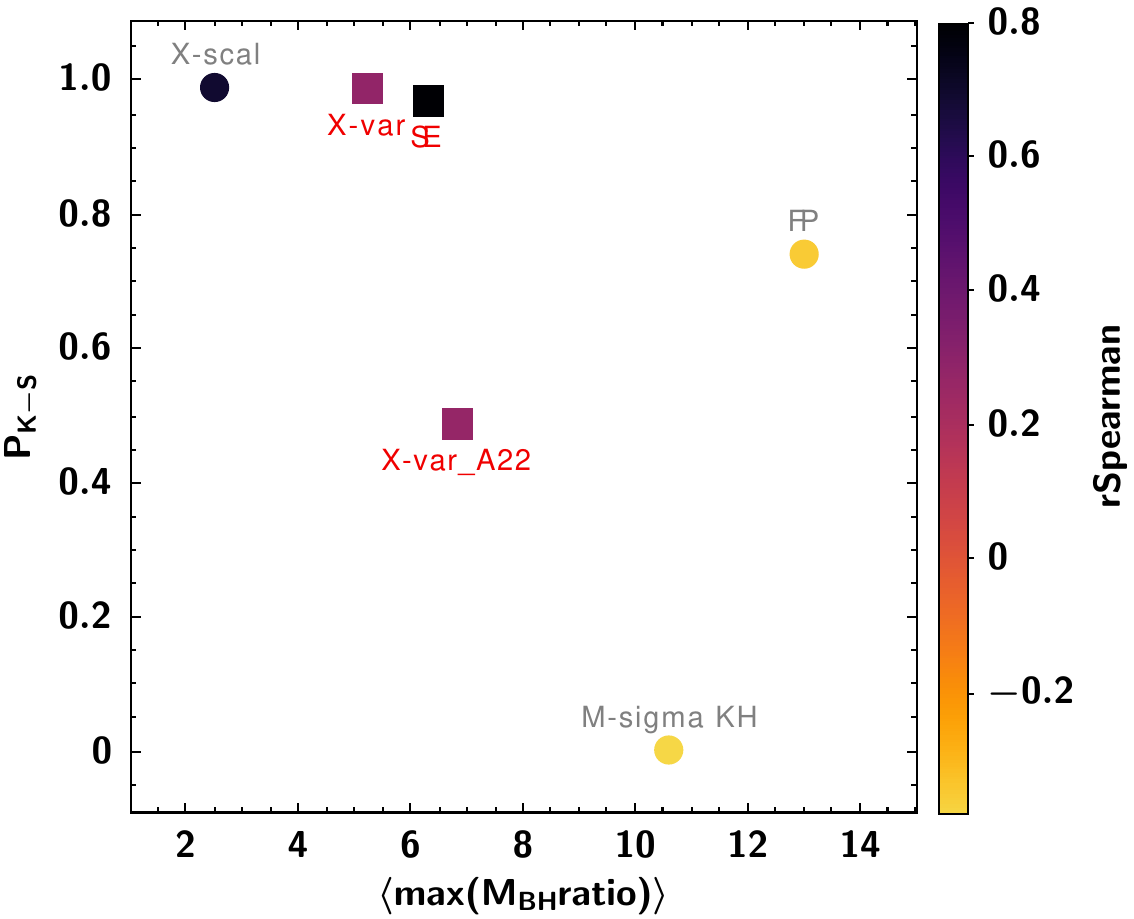}
\caption{The quantity along the x-axis is $\langle {\rm max}(M_{\rm BH} {\rm ratio}) \rangle$, the average of the maximum between $M_{\rm BH,dyn}/M_{\rm BH,indirect}$ and $M_{\rm BH,indirect}/M_{\rm BH,dyn}$, computed for each source: the smaller the x value the better the agreement with the dynamical value.
The quantity along the y-axis is the probability associated with the K--S test: higher values indicate that the distribution of the \mbh\ obtained with an indirect method is consistent with that of dynamical values.
Finally, the auxiliary color-coded axis represents the Spearman's rank correlation coefficient $r$: darker colors indicate a positive correlation, light colors indicate a negative or nonexistent linear correlation between \mbh\ indirect values and the corresponding dynamical ones. Circles represent methods that were applied to the restricted dynamical sample, whereas squares refer to the indirect methods applied to the extended dynamical sample.
Dark symbols in the top left corner indicate the best agreement between indirect and dynamical methods; the opposite is true for the bottom right corner and lighter colors.}
\label{fig:KSvsMBHratio}
\end{figure}

\subsection{Comparison with the dynamical sample}
A simple way to test the reliability of indirect methods is to compare their \mbh\ distributions with that derived using dynamical methods. This is illustrated in Fig.~\ref{fig:histog}, where the histograms yielded by the \msigma, FP, and X-ray scaling methods are compared to the histogram of the restricted dynamical sample (top panels), and the distributions of \mbh\ produced by the SE and the X-ray variability method (using both the \citet{Akylas2022} prescription and the correlation derived in this work) are compared to the histogram of the extended dynamical sample (bottom panels).

From this figure, one can conclude that there is a reasonably good overlap between the dynamically based \mbh\ distributions and the distributions obtained with most of the indirect methods tested here, with the notable exception of the distribution derived from the \msigma\ correlation, which appears to be shifted to the right toward larger values (top left panel).

In order to quantify the agreement (or lack thereof) between the \mbh\ distributions obtained with the various indirect methods and the one inferred from dynamical measurements, we carried out Kolgomorov--Smirnov (K--S) and Student's $t$ tests, which compare the whole distributions and their means, respectively. Large values of the probabilities associated with these tests indicate that the distributions (and means) are statistically consistent with the dynamical values. The results of these tests are reported in the first four columns of Table \ref{tab:indirect_methods} and are in general agreement with the qualitative conclusions inferred from the visual inspection of Fig.~\ref{fig:histog}.

Although comparing the distributions of \mbh\ offers a simple way to reveal whether an indirect method is inconsistent with dynamical measurements (as in the case of the \msigma\ method), it may be misleading in suggesting that other methods are fully consistent with the dynamical estimates. For example, if the same indirect method happens to yield for some objects \mbh\ measurements that are substantially larger and for others substantially smaller than the accepted dynamical ones, the overall distribution may appear to be consistent with the dynamical one (as in the case of the FP method). In other words, the formal consistency between \mbh\ distributions should be considered as a necessary but not sufficient condition for an indirect method to be considered reliable.

A more direct way to illustrate whether an indirect method is accurate is to plot the \mbh\ obtained from indirect methods versus the corresponding dynamical ones, as shown in Fig.~\ref{fig:xy_dyn_indir}, where the continuous line indicates the perfect match $M_{\rm BH,indirect}/M_{\rm BH,dyn} = 1$ and the dashed lines represent ratios of 3 and 1/3, which are of the order of the typical uncertainties associated to \mbh\ estimates. This figure qualitatively confirms that there is a good agreement between dynamically estimated \mbh\ and values obtained with the X-ray scaling method (top right panel), the SE method (bottom left panel), as well as with the values obtained with the X-ray normalized excess variance (bottom middle and right panels). In the latter case, for completeness, we have shown both the values obtained with the best fit correlation obtained by \citet{Akylas2022} using a larger RM sample (bottom middle panel) and
the best fit obtained in this work using the extended dynamical sample (bottom right panel). Both choices provide values in general agreement with the dynamical ones.

Fig.~\ref{fig:xy_dyn_indir} also confirms that a substantial number of \mbh\ values derived with the \msigma\ correlation method are overestimated compared to the corresponding dynamical ones (top left panel) and indicates that in some cases the FP estimates may exceed the dynamical values by orders of magnitude and in others yield significantly lower values (top middle panel).

A quantitative way to assess the agreement (or lack thereof) illustrated in Fig.~\ref{fig:xy_dyn_indir} is based on the Spearman's and Kendall's rank correlation coefficients $r$ and $\tau$ and their associated probabilities, with a positive correlation expected for the indirect methods that yield \mbh\ estimates consistent with the dynamical ones. The results of these statistical tests are reported in the fifth and sixth columns of Table \ref{tab:indirect_methods} and corroborate the previous conclusions with negative correlation coefficients obtained for the \msigma\ and FP methods and positive correlation coefficients for the SE and the two X-ray based methods.

An additional straightforward way to directly compare the \mbh\ estimates from indirect methods with the corresponding dynamical ones is obtained by computing their ratio. However, since the same method can yield for some objects estimates that are much larger and for others much smaller than the dynamical ones, the differences may cancel out when computing the average of the ratios for the whole sample. To circumvent this problem, for each source we computed the maximum value between $M_{\rm BH,dyn}/M_{\rm BH,indirect}$ and $M_{\rm BH,indirect}/M_{\rm BH,dyn}$ (that is, a value that by definition cannot be smaller than 1) and then calculated its mean,
$\langle {\rm max}(M_{\rm BH}\, {\rm ratio}) \rangle$, for the dynamical samples. The results of this analysis, reported in the last column of Table \ref{tab:indirect_methods}, confirm our previous conclusions with smaller values of  $\langle {\rm max}(M_{\rm BH}\, {\rm ratio}) \rangle$ associated with the X-ray scaling method, the X-ray variability one, and the SE.

A synthesis of the various statistical comparisons of indirect methods with the dynamical ones is illustrated in Fig.~\ref{fig:KSvsMBHratio}, where circles indicate indirect methods tested using the restricted dynamical sample and squares refer to methods tested using the extended dynamical sample. In this diagram, the x-axis represents $\langle {\rm max}(M_{\rm BH}\, {\rm ratio}) \rangle$ and the y-axis the probability associated with the K--S test,
whereas the auxiliary color-coded axis indicates the Spearman's correlation rank coefficient $r$. Based on these criteria, darker symbols located in the top left corner imply a good agreement between indirect and dynamical methods; the opposite is true of the lighter-colored symbols located in the bottom right corner. 

Based on this diagram, we can conclude that the X-ray scaling method shows the best agreement with the dynamical measurements according to the three criteria illustrated in Fig.~\ref{fig:KSvsMBHratio} (the symbol is dark, indicating the positive correlation, and located in the top left corner). Also the SE method (which is only applicable to AGN with broad lines) shows a good agreement, as well as the X-ray variability method. In the latter case, Fig.~\ref{fig:KSvsMBHratio} reveals a clear distinction between the method based on the correlation obtained by \citet{Akylas2022}, which shows a moderately good agreement with the dynamical values, and the one obtained in this work, which is more consistent with the dynamical one. Finally, the same figure reinforces the conclusion that neither
the \msigma\ method nor the fundamental plane of BH activity yields \mbh\ measurements consistent with the dynamical values. 
\begin{figure*}
\includegraphics[width=0.98\columnwidth]{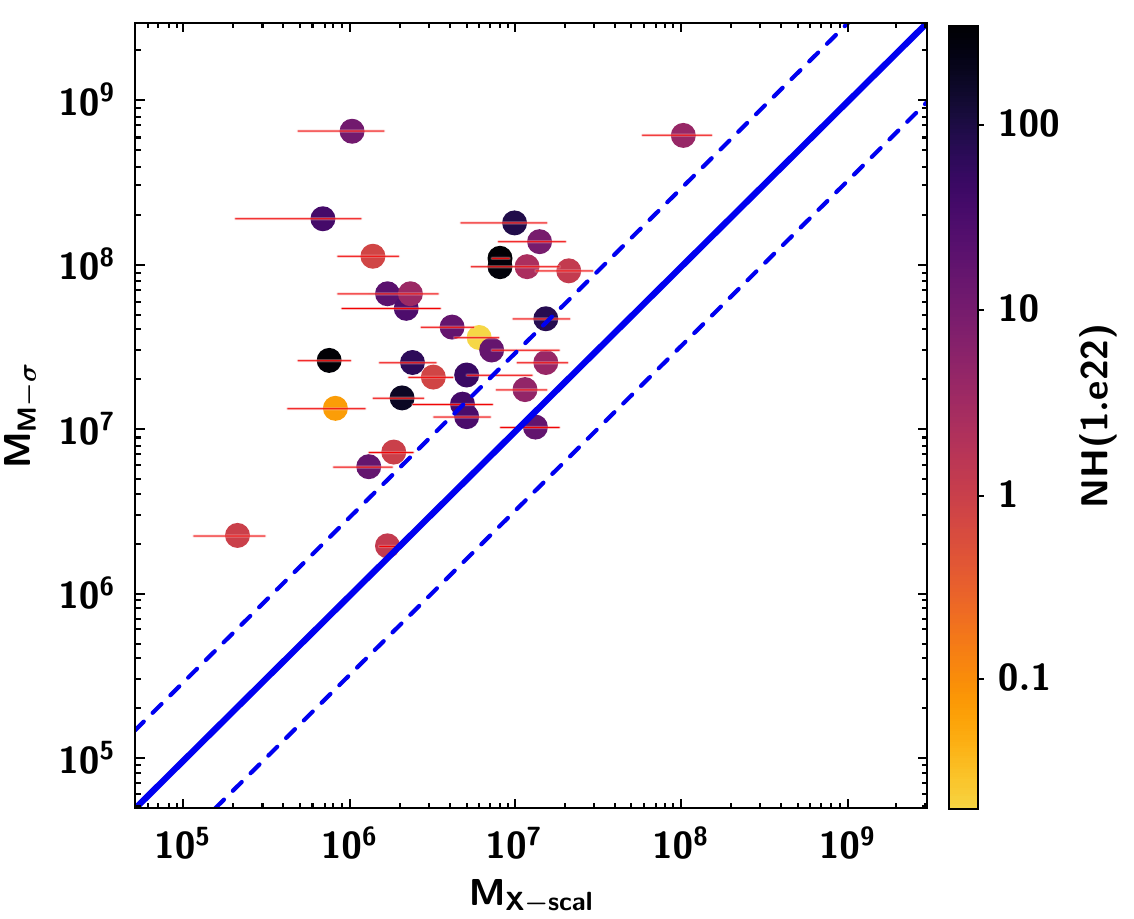} \includegraphics[width=0.98\columnwidth]{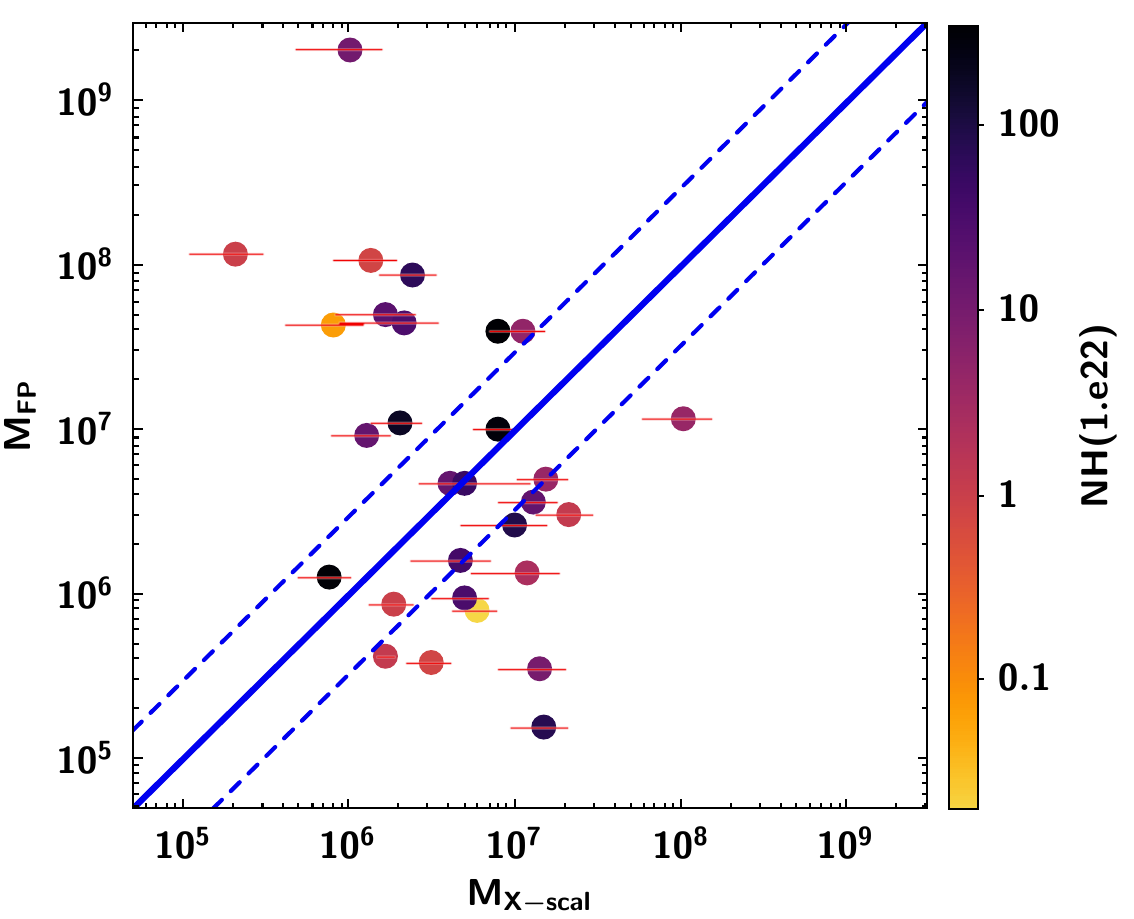}
\includegraphics[width=0.98\columnwidth]{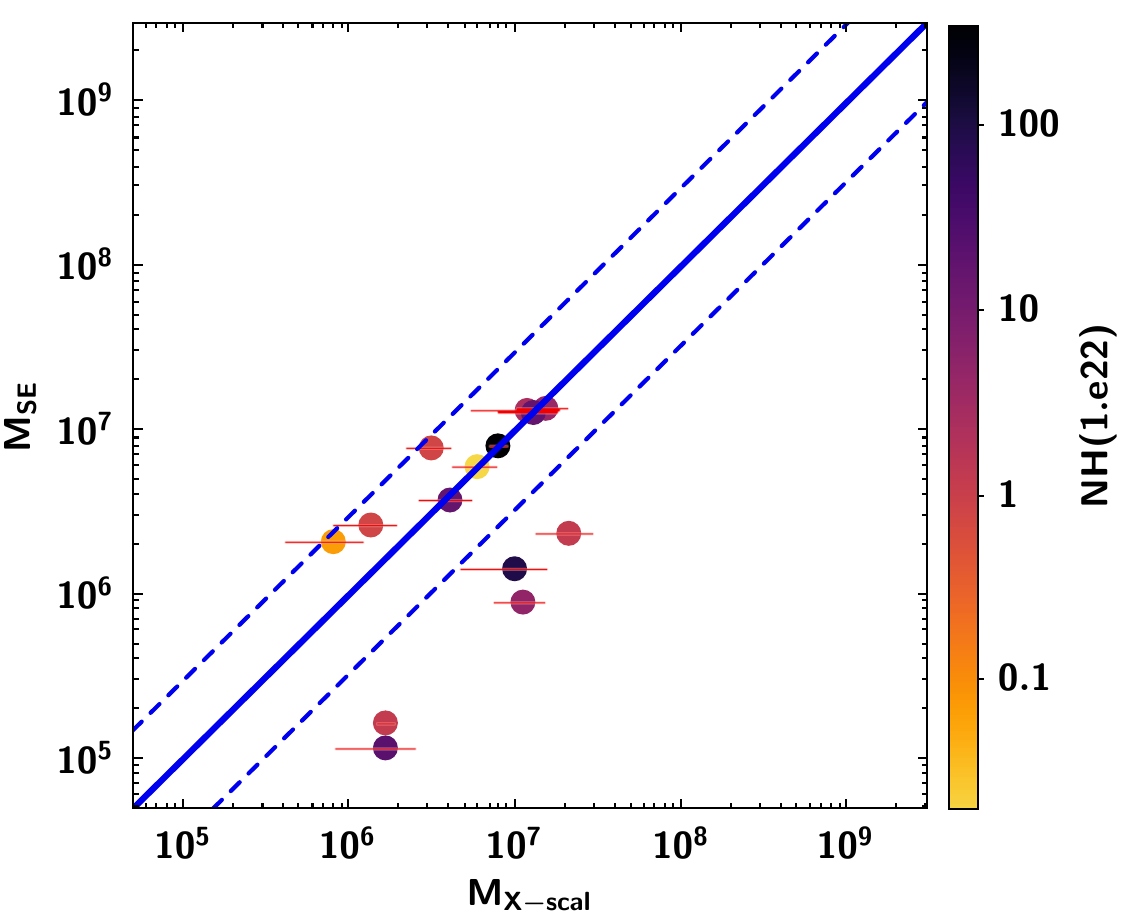} \includegraphics[width=0.98\columnwidth]{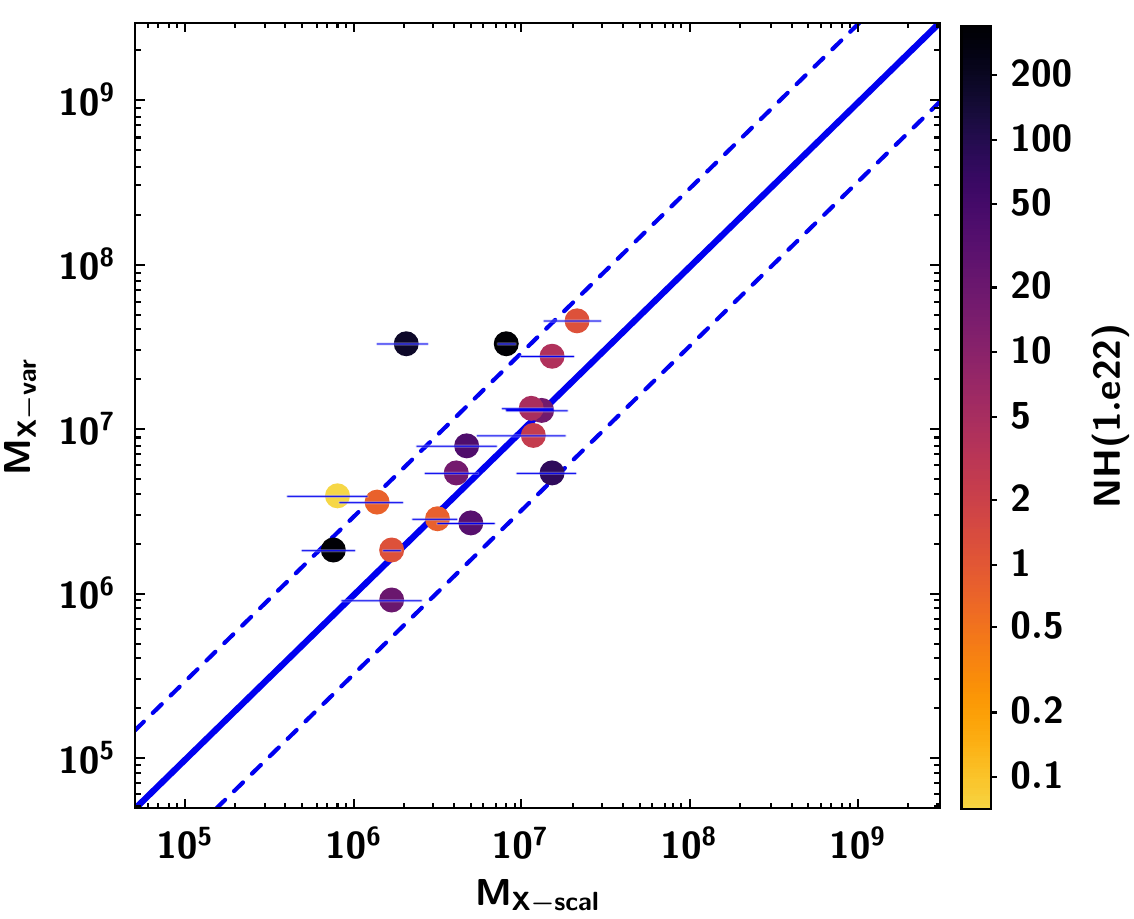}
\caption{Top left panel: \mbh\ obtained with the \msigma\ correlation plotted vs. those from the X-ray scaling method. Top right panel: \mbh\ values from the fundamental plane of black hole activity vs. those from the X-ray scaling method. Bottom left panel: \mbh\ obtained with the SE method plotted vs. the values derived from the X-ray scaling method.
Bottom right panel: \mbh\ from X-ray variability of 80 ks segments vs. those from the X-ray scaling method. The symbols are color coded based on the intrinsic \nh\ along the line of sight. 
}
\label{fig:xy_Xscal_indir}
\end{figure*}

\begin{figure}
\includegraphics[width=\columnwidth]{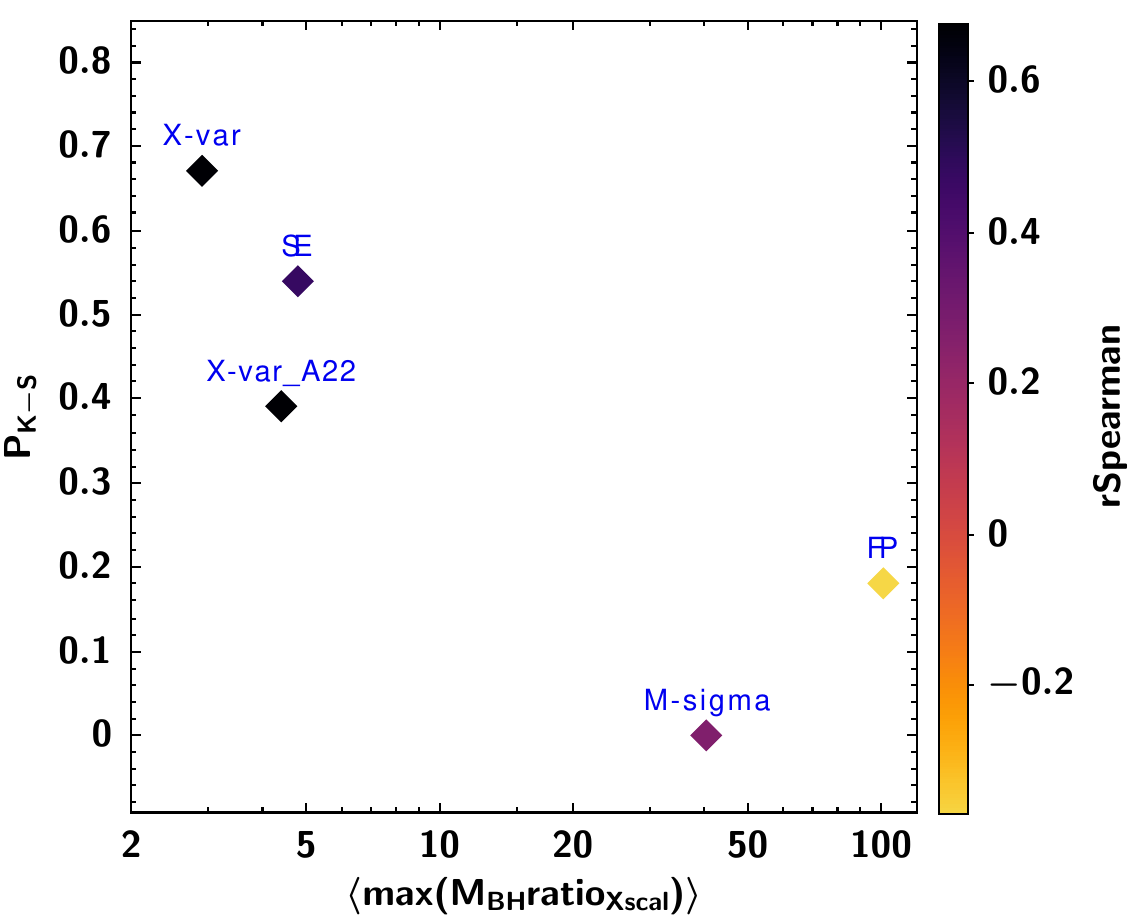}
\caption{Probability associated with the K--S test plotted vs.
$\langle {\rm max}(M_{\rm BH} {\rm ratio_{Xscal}}) \rangle$, the average of the maximum between $M_{\rm BH,Xscal}/M_{\rm BH,indirect}$ and $M_{\rm BH,indirect}/M_{\rm BH,Xscal}$. The auxiliary color-coded axis represents the Spearman's rank coefficient $r$.  Dark symbols in the top left corner indicate good agreement between X-ray scaling methods and the other indirect methods; the opposite is true for the bottom right corner.}
\label{fig:KSvsMBHratio_Xscal}
\end{figure}
\subsection{Comparison for the whole sample}
According to all statistical tests performed so far, the X-ray scaling method appears to yield \mbh\ estimates that are consistently in good agreement with the dynamical values. We will therefore consider this method as an accurate proxy of the dynamical \mbh\ estimate and carry out a similar comparative analysis for the whole sample of AGN. The direct comparison with \msigma, FP, SE, and X-ray excess variance methods is shown in Fig.~\ref{fig:xy_Xscal_indir}, where the \mbh\ estimates obtained with these four indirect methods are plotted versus the values from the X-ray scaling method. Once more the \msigma\ method confirms the tendency to systematically overestimate the \mbh, whereas FP often yields values that differ by orders of magnitude in both directions from those considered more reliable. The SE method shows a generally good agreement with the exception of a handful of objects whose mass appears to be underestimated by this method. Finally, it is interesting to note the surprisingly good agreement between the two X-ray based methods, which estimate \mbh\ starting from very different assumptions. In Fig.~\ref{fig:KSvsMBHratio_Xscal} we illustrate the results of this comparison showing the diagram $P_{\rm K-S}$ vs. $\langle {\rm max}(M_{\rm BH}\, {\rm ratio}_{\rm Xscal}) \rangle$, which is defined as the average of the maximum between $M_{\rm BH,Xscal}/M_{\rm BH,indirect}$ and $M_{\rm BH,indirect}/M_{\rm BH,Xscal}$. Based on this figure we conclude that, using all the available sources in this volume-limited sample, the X-ray variability method, based on the correlation determined in this work, shows the best agreement with the X-ray scaling method. A reasonably good agreement is achieved also by the SE method and the X-ray variability one based on the \citet{Akylas2022} correlation. Conversely, the \msigma\ and FP methods appear to be inconsistent with the X-ray scaling method.


\section{Discussion}
\label{sec:discussion}
The primary goal of this work is to compare different indirect methods to determine the \mbh\ in AGN, with particular emphasis on methods that could be applied to heavily obscured objects, where the most frequently used optically based methods are not applicable. To this end, we have used a volume-limited sample of 32 AGN (D < 40 Mpc) selected from the 70-month BAT catalog, which can be considered representative of the AGN population in the local universe, since the detection at very hard X-rays strongly reduces the bias towards unobscured AGN. We have taken advantage of the second data release of the Swift BAT AGN Spectroscopic Survey (BASS DR2), which provides a wealth of useful information, including reliable redshift-independent distances, which play a crucial role in the determination of \mbh\ in some methods, including the direct dynamical ones.

We started from two subsamples of 11 (the restricted dynamical sample) and 17 AGN (the extended dynamical sample) respectively, with \mbh\ determined via dynamical methods to assess the reliability of different indirect methods applicable to all AGN regardless of their level of obscuration, such as the X-ray scaling method, the X-ray variability one based on the normalized excess variance, the \msigma\ correlation, and the fundamental plane of BH activity. The first important result is that the \mbh\ values derived from X-ray based methods are in good agreement with the dynamical values. On the contrary, the values obtained with the \msigma\ correlation are systematically and substantially overestimated with respect to the dynamical values, whereas the FP method often yields values that differ from the dynamical ones by one or two order of magnitudes in both directions (see Fig.~\ref{fig:xy_dyn_indir}).

\subsection{X-ray based methods}
The bottom right panel of Fig.~\ref{fig:xy_Xscal_indir} illustrates the remarkable agreement between the X-ray scaling method and the X-ray variability one. Note that the \mbh\ values derived from the normalized excess variance method are based on a best fit anti-correlation, $\log M_{\rm BH}=-1.04\times\log(\sigma^2_{\rm NXV}/0.005)+7.14$, obtained from the extended dynamical sample, which by construction will yield values fully consistent with the dynamical ones. This anti-correlation is slightly different from that obtained by \citet{Akylas2022}, based on a large sample of AGN with RM masses,  $\log M_{\rm BH}=-1.13\times\log(\sigma^2_{\rm NXV}/0.005)+7.38$, which still yields \mbh\ in broad agreement with the dynamical values but at a lower statistical level (see Figs. \ref{fig:KSvsMBHratio} and \ref{fig:KSvsMBHratio_Xscal}).

These two X-ray methods are based on completely different assumptions and methodologies. More specifically, the X-ray scaling method scales up the dynamical mass of a stellar-mass BH under the assumptions that X-rays in XRBs and AGN are produced by the same Comptonization process and that the photon index $\Gamma$ is a reliable indicator of the accretion rate in Eddington units. On the other hand, the method based on the \xsvar\ is completely model-independent and exploits the universal anti-correlation observed in AGN between \mbh\ and X-ray variability. Also their respective limitations are different: for example, the X-ray scaling method by construction cannot be applied to very low-accreting BHs because in that regime $\Gamma$ is not positively correlated with \laedd\ \citep{Jang2014}, but for moderate and highly accreting objects this method constrains \mbh\ with uncertainties ranging between 15 per cent and 40 per cent depending on the quality of X-ray spectra and the reference source utilized (see, e.g. \citealt{Williams2023}). The X-ray variability method instead has no specific limitations for very low-accreting objects; however, being a statistical method, it yields \mbh\ values with an intrinsic uncertainty of the order of 0.25 dex (the average scatter around the best fit) and can be currently applied to a fairly limited number of AGN since objects with sufficiently long, good-quality hard X-ray light curves are still a minority compared to those with good-quality spectra.
Considering that the X-ray variability method by construction must yield values in agreement with the dynamical ones, the full consistency with the X-ray scaling method confirms that \mbh\ estimates derived with the latter method can be considered as reliable surrogates of dynamical values. We therefore utilize the values obtained with the X-ray scaling method as a reference for the comparison with those from other indirect methods over the entire sample. 

\subsection{SE method}
Although we are mainly interested in constraining \mbh\ in heavily obscured AGN, we can use the broad-lined AGN contained in our volume-limited sample (14 objects out of 32) and their X-ray scaling \mbh\ estimates to assess the reliability of the SE method, which is the most utilized indirect method to determine  \mbh\ in type 1 AGN. This test is illustrated in the bottom left panel of Fig.~\ref{fig:xy_Xscal_indir}, where the \mbh\ obtained with the SE technique are plotted vs. those from the X-ray scaling method. This figure suggests a reasonably good agreement between the two methods, which is confirmed by the same statistical tests carried out systematically throughout this work: $P_{\rm K-S}=0.54$, $P_{\rm t}=0.24$,  $r=0.48$ ($P_{\rm S}=0.08$), $\tau=0.38$ ($P_{\rm K}=0.06$), and $\langle {\rm max}(M_{\rm BH} {\rm ratio}) \rangle =4.8$ and are summarized in Fig.~\ref{fig:KSvsMBHratio_Xscal}. 

However, the same figure also reveals a handful of objects apparently underestimated by the SE by nearly one order of magnitude with respect to the corresponding X-ray scaling values. Based on our X-ray spectral analysis, all these outliers possess \nh\ along the line of sight in excess of $2\times10^{22}~{\rm cm^{-2}}$ with three of them (NGC 2992, NGC 1365, and NGC 5728) of the order of a few units in $10^{23}~{\rm cm^{-2}}$ and one (NGC 5506) in the CT regime, suggesting that the discrepancy may be ascribed to the partial obscuration of the BLR that reduces the observed line width leading to \mbh\ that are underestimated. 

This result appears to be in qualitative agreement with the conclusions of \citet{Caglar2020} and \citet{Mejia-Restrepo2022}, who suggest the need of a correction to the \mbh\ values obtained with the SE in heavily absorbed AGN. However, their conclusions were based on the comparison with \mbh\ from the \msigma\ correlation, under the assumption that the latter yields reliable values, which is at odds with the findings of this work. The bottom left panel of Fig.~\ref{fig:xy_Xscal_indir} also shows that other sources that are heavily absorbed yield SE-based \mbh\ that are fully consistent with the X-ray scaling estimates. This may suggest that there is not a simple corrective factor or a monotonic function of \nh\ that can be systematically applied to all SE estimates in heavily obscured AGN.

\begin{figure}
\includegraphics[width=\columnwidth]{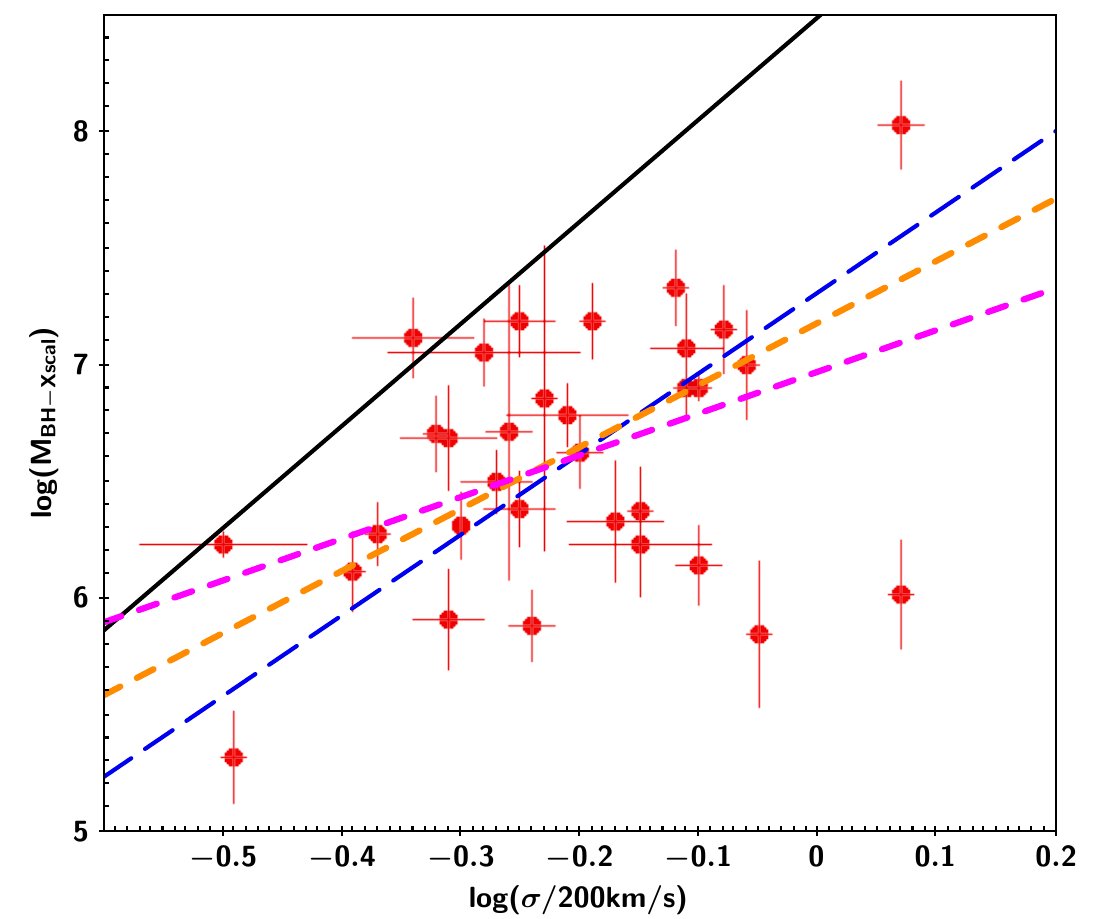}
\caption{\mbh\ obtained with the X-ray scaling method for the whole sample plotted vs. the bulge stellar velocity dispersion $\sigma_\star$ divided by 200 km s$^{-1}$. The black continuous line represents the \msigma\ correlation of \citet{Kormendy2013}, the blue long-dashed line indicates the correlation of \citet{Woo2013}, the magenta short-dashed line illustrates the best-fit correlation, $\log \left( \frac{M_{\mathrm{BH}}}{\mathrm{M}_\odot} \right) = 6.97 + 1.79 \log \left( \frac{\sigma_\star}{200\, \mathrm{km\, s}^{-1}} \right)$ derived in this work using the linear regression routine \textsc{linmix\textunderscore err}, whereas the orange short-dashed line represents the best-fit correlation, $\log \left( \frac{M_{\mathrm{BH}}}{\mathrm{M}_\odot} \right) = 7.18 + 2.66 \log \left( \frac{\sigma_\star}{200\, \mathrm{km\, s}^{-1}} \right)$ derived in this work using the linear least-square routine \textsc{fitexy}.}
\label{fig:xy_Mxssigma}
\end{figure}

\subsection{M-sigma correlation}
When good-quality X-ray data exist (that is, high signal-to-noise broadband spectra or light curves with sufficiently long exposures), then X-ray based methods appear to be the  most reliable way to constrain \mbh. Unfortunately, the current paucity of high-quality X-ray data limits the application of these methods to a few hundreds of AGN. On the other hand, since reliable optical data of host galaxy bulges exist for a much larger number of AGN, it is important to investigate the possibility of applying a correction factor to make the \mbh\ obtained with the \msigma\ correlation more consistent with those from the X-ray scaling method (and hence with the dynamical ones). The top panel of Fig.~\ref{fig:xy_Xscal_indir}, showing that the values obtained with the \msigma\ correlation are systematically overestimated compared to the X-ray scaling estimates for the whole sample, confirms and reinforces the conclusion derived from the comparison with the dynamical \mbh\ estimates. This conclusion appears to be in agreement with the findings of \citet{Greene2016} and \citet{Ricci2017c}, who compared the \mbh\ estimates from the \msigma\ correlation of heavily obscured AGN to the values obtained either from maser measurements or from an IR-based SE method, and found that the \msigma\ values are substantially overestimated.

The reason why in this work we utilized the \msigma\ correlation of \citet{Kormendy2013} is twofold: 1) \citet{Koss2022b} utilize the correlation of \citet{Kormendy2013} to estimate the \mbh\ for several hundreds of AGN in the BASS DR2. Therefore, determining a suitable correction factor would provide the simplest way to quickly obtain more reliable \mbh\ (and, consequently, the Eddington ratio \laedd) values for hundreds of AGN. 2) Based on our previous work, where we carried out a systematic comparison of several indirect methods using NGC 4151 as a reference, only the correlation of \citet{Kormendy2013} yielded an estimate consistent with the \mbh\ obtained from different dynamical methods \citep{Williams2023}.

To investigate whether a different \msigma\ correlation may provide better estimates of \mbh\ in our volume-limited sample, in Fig.~\ref{fig:xy_Mxssigma} we plotted the \mbh\ obtained with the X-ray scaling method versus the bulge stellar velocity dispersion $\sigma_\star$ and overlaid the \msigma\ correlation of \citet{Kormendy2013} (black continuous line), as well as the correlation of \citet{Woo2013} (the blue long-dashed line). 

For completeness, we also independently computed the best linear fit with three different methods: we first used \textsc{ladfit}, a robust least absolute deviation method, which does not take into account the errors but is unaffected by outliers, and obtained $\log \left( \frac{M_{\mathrm{BH}}}{\mathrm{M}_\odot} \right) = (7.14) + (2.36)\times \log \left( \frac{\sigma_\star}{200\, \mathrm{km\, s}^{-1}} \right)$. Then, we utilized the linear regression routine \textsc{linmix\textunderscore err} that accounts for the uncertainties on both variables using a Bayesian approach \citep{Kelly2007}, obtaining the linear regression $\log \left( \frac{M_{\mathrm{BH}}}{\mathrm{M}_\odot} \right) = (6.97\pm0.03) + (1.79\pm0.13)\times \log \left( \frac{\sigma_\star}{200\, \mathrm{km\, s}^{-1}} \right)$ with a standard deviation of 0.75 on the posterior distribution of the slope values (this fit is represented by the short-dashed magenta line in Fig.~\ref{fig:xy_Mxssigma}). Finally, using \textsc{fitexy}, another linear least-squares routine that accounts for the the uncertainties on both x and y, we obtained $\log \left( \frac{M_{\mathrm{BH}}}{\mathrm{M}_\odot} \right) = (7.18\pm0.06) + (2.66\pm0.27)\times \log \left( \frac{\sigma_\star}{200\, \mathrm{km\, s}^{-1}} \right)$ (represented by the short-dashed orange line in Fig.~\ref{fig:xy_Mxssigma}). 

Regardless of the routine used for the linear fit, the slope obtained is fairly shallow, which is marginally consistent with the correlation of \citet{Woo2013} but not with that of \citet{Kormendy2013}. In the remainder of the paper, for our independently derived correlation we will use the results from  \textsc{fitexy}, which accounts for both uncertainties in x and y, is fully consistent with the results from \textsc{ladfit}, and has a slope slightly steeper than that derived with \textsc{linmix\textunderscore err}. However, the same conclusions can be derived from the latter linear fit as well. 

Importantly, we note that our results, based on a volume-limited unbiased sample (D < 40 Mpc which roughly corresponds to z < 0.01) of type 1 and type 2 AGN selected from the BAT survey, are in full agreement with those recently obtained by \citet{Caglar2023}, where an \msigma\ correlation with a slope of $3.09\pm0.39$ was derived using a larger volume limited sample (z < 0.08) of 154 type 1 AGN from the 105-month BAT sample with reliable \mbh\ derived from the SE method. This may suggest that the \msigma\ correlations derived from inactive or nearly quiescent galaxies, such that of \citet{Kormendy2013}, are not appropriate to determine the \mbh\ in AGN.

\begin{figure}
\includegraphics[width=\columnwidth]{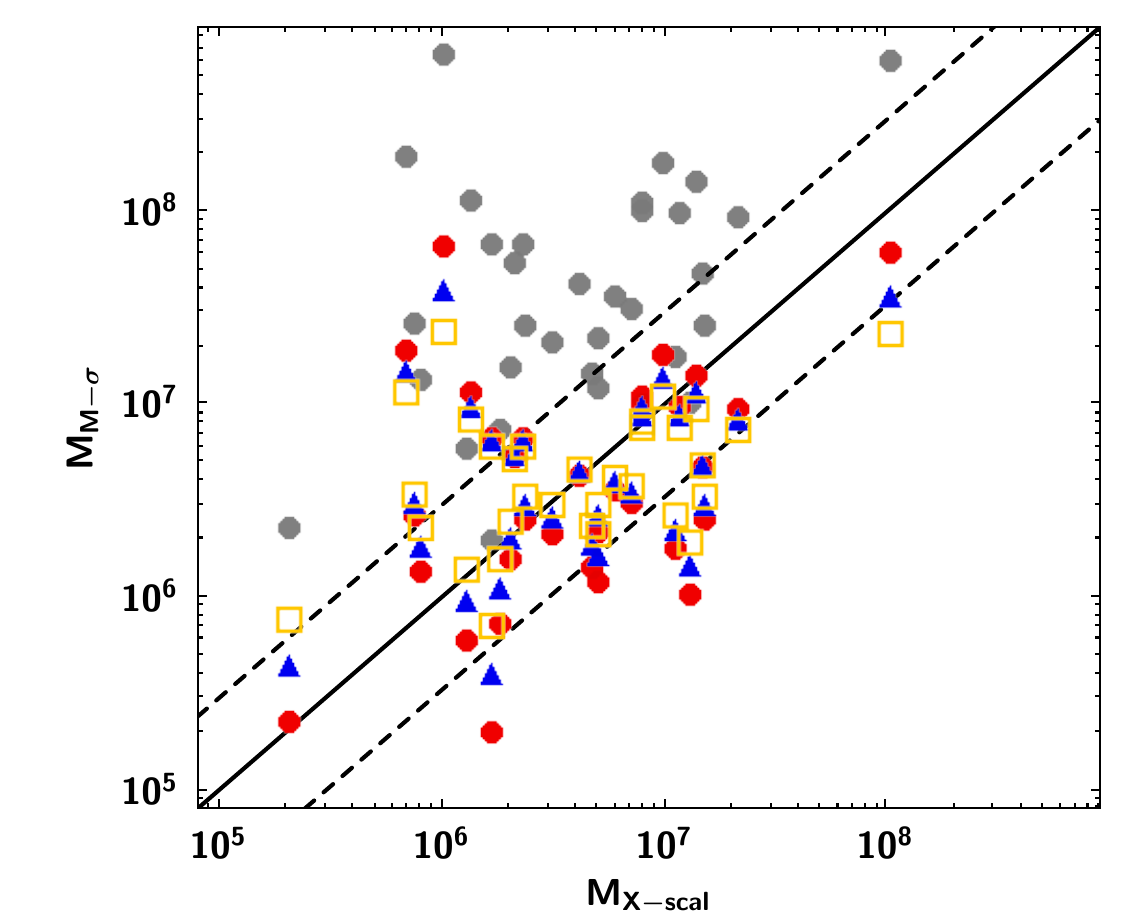}
\caption{\mbh\ obtained with different \msigma\ correlations plotted vs. those from the X-ray scaling method. The grey circles indicate the values obtained from the original correlation from \citet{Kormendy2013}, whereas the red circles show the same values scaled down by a factor of 10. The blue triangles represent the \mbh\ derived from the correlation of \citet{Woo2013}, and the open orange squares are the values obtained from the \msigma\ correlation derived in this work.}
\label{fig:xy_MsigallvsXscal}
\end{figure}
 From Fig.~\ref{fig:xy_Mxssigma} it is clear that the \citet{Woo2013} \msigma\ correlation, as well as the ones derived in this work, provide a better fit of the data while the correlation from \citet{Kormendy2013} overestimates most of the \mbh. Since the discrepancy appears to be related mostly to the intercept, in principle one could still obtain a reasonable estimate of \mbh\ simply renormalizing the \citet{Kormendy2013} correlation. This process is illustrated in Fig.~\ref{fig:xy_MsigallvsXscal}, which shows that the \mbh\ estimates, obtained from the \msigma\ correlation of \citet{Kormendy2013} (grey circles), become broadly consistent with the corresponding X-ray scaling values once they are scaled down by a factor of 10 (red circles), and largely overlap the estimates derived from the \msigma\ correlation of \citet{Woo2013} (illustrated by the blue triangles), and the ones obtained from our best-fitting model (orange open squares).

\begin{figure}
\includegraphics[width=\columnwidth]{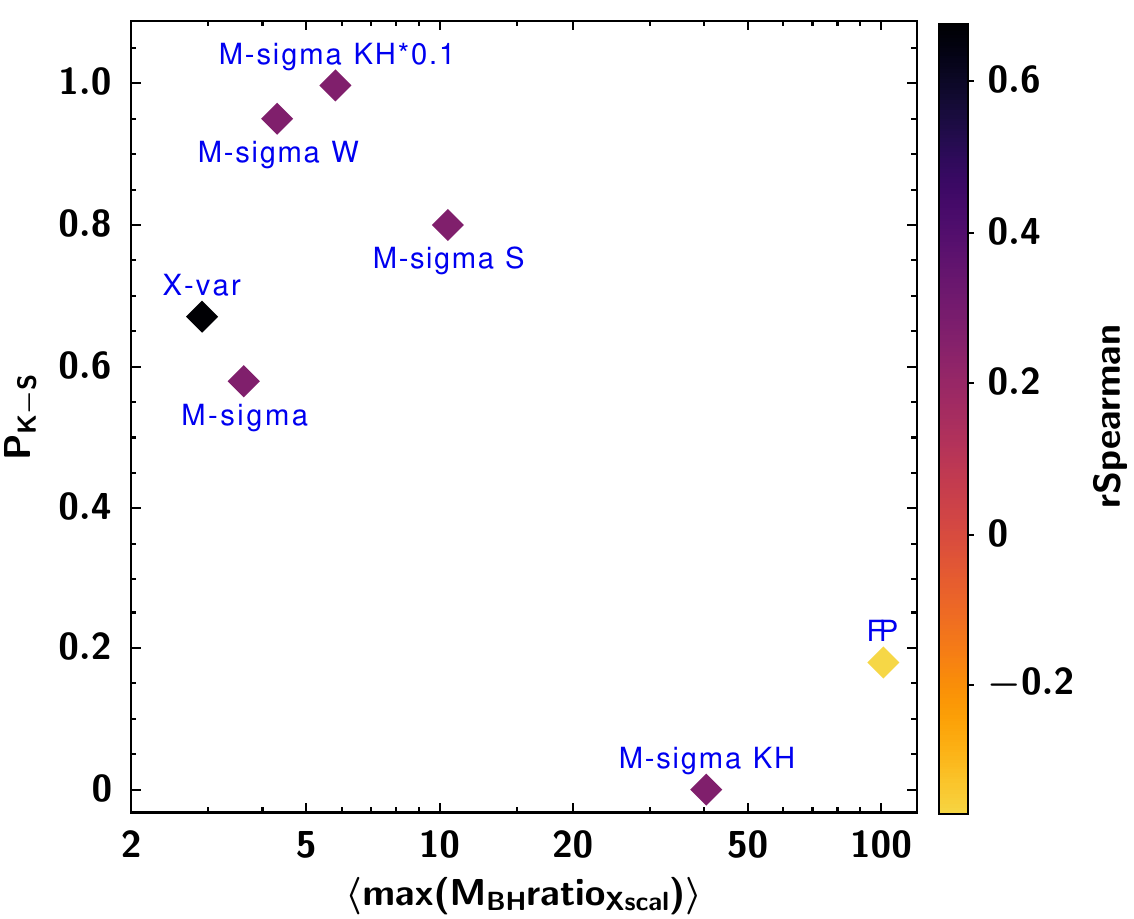}
\caption{Probability associated with the K--S test plotted vs.
$\langle {\rm max}(M_{\rm BH} {\rm ratio}) \rangle$, the average of the maximum between $M_{\rm BH,X-scal}/M_{\rm BH,indirect}$ and $M_{\rm BH,indirect}/M_{\rm BH,X-scal}$ using the whole sample.
The auxiliary color-coded axis represents the Spearman's rank coefficient $r$.
In this plot we show the different locations in the diagram of the various \msigma\ correlations, along with the location of the X-ray variability and FP values. {\it M-sigma W} indicates the correlation of \citet{Woo2013},  {\it M-sigma KH} the correlation of \citet{Kormendy2013}, and {\it M-sigma KH*0.1 } the same correlation scaled down by a factor of 10; {\it M-sigma S} represents the correlation of \citet{Shankar2016}; finally, {\it M-sigma} is the one obtained in this work.}
\label{fig:KSvsMBHratio_Msig}
\end{figure}

A more comprehensive comparison of the various \msigma\ correlations is shown in Fig.~\ref{fig:KSvsMBHratio_Msig}, where we utilize the same diagnostic diagram introduced in Section \ref{sec:comparison} to illustrate the statistical comparisons of the different indirect methods. In this figure,  {\it M-sigma} represents the correlation derived in this work, {\it M-sigma W} indicates the correlation of \citet{Woo2013},  {\it M-sigma KH} the correlation of \citet{Kormendy2013}, and {\it M-sigma KH*0.1 } the same correlation scaled down by a factor of 10; finally, {\it M-sigma S} represents the correlation of \citet{Shankar2016}.  Fig.~\ref{fig:KSvsMBHratio_Msig} reveals that \mbh\ estimates obtained with the correlation of \citet{Kormendy2013} scaled down by a factor of 10 (which yields values similar to those obtained from \citet{Woo2013}) have a considerably better agreement with the \mbh\ values from the X-ray scaling method (and hence the dynamical ones) than the original \citet{Kormendy2013} correlation. This figure also reveals that, based on the $\langle {\rm max}(M_{\rm BH} {\rm ratio}) \rangle$ value, which is the most stringent of the three criteria depicted in the diagram, the \msigma\ correlation derived in this work shows the best agreement with the accepted \mbh\ estimates.


\section{Conclusion}
Here, we summarize our main results, obtained using a volume-limited sample of AGN selected from the 70-month BAT catalog:\\
\begin{itemize}
\item Of all indirect methods used to estimate \mbh\ in obscured AGN, which were tested in this work, only the X-ray based ones (the scaling method and the variability one based on the normalized excess variance) yield values that are consistently in agreement with the dynamical estimates. For this reason, in the absence of a more direct dynamical way to constrain \mbh\ and when good-quality X-ray data with sufficiently long exposure are available, X-ray based estimates should be considered as the most reliable and used as a reference to calibrate other indirect methods.\\

\item BH values obtained with the \msigma\ correlation proposed by \citet{Kormendy2013}, which was also used to estimate the vast majority of \mbh\ in the BASS DR2 \citep{Koss2022b}, appear to be systematically overestimated when compared to dynamical measurements (and hence to X-ray based ones). Scaling down these estimates by a factor of 10 substantially improves the agreement with the corresponding dynamical values (and X-ray based estimates) and represents the fastest way to obtain reasonable \mbh\ for hundreds of AGN. \\

\item In general, AGN \mbh\ values appear to be more consistent with the estimates derived from the \msigma\ correlation obtained by \citet{Woo2013} for an RM AGN sample or the one derived in this work, $\log \left( \frac{M_{\mathrm{BH}}}{\mathrm{M}_\odot} \right) = (7.18\pm0.06) + (2.66\pm0.27)\times \log \left( \frac{\sigma_\star}{200\, \mathrm{km\, s}^{-1}} \right)$. These correlations with shallower slopes are consistent with the one recently derived by \citet{Caglar2023} using a larger unbiased volume-limited sample of type 1 AGN. This may suggest that AGN follow a different \msigma\ correlation from the one inferred using inactive or nearly quiescent nearby galaxies.\\

\item For the broad-lined AGN of our sample, a statistical comparison between SE and X-ray scaling estimates of \mbh\ indicates a good agreement, but also confirms the tendency of the SE technique to underestimate the \mbh\ in sources affected by substantial obscuration. It is therefore important to properly correct the SE values of heavily obscured sources, as suggested by \citet{Mejia-Restrepo2022}. However, the corrections should not be based on the comparison with the \msigma\ correlation from inactive galaxies, since the latter has a tendency to overestimate the \mbh. Our analysis also reveals that some heavily absorbed sources yield SE-based \mbh\ that are fully consistent with the X-ray scaling estimates, suggesting that there is not a simple corrective factor that can be uniformly applied to all SE estimates in heavily obscured AGN.\\

\item As expected from previous studies, the \mbh\ estimates obtained with the FP are fairly inconsistent and unreliable, with values that can differ from the dynamical ones by two orders of magnitude in both directions and with only a small fraction of objects with \mbh\ consistent with dynamical values (and X-ray based ones). Therefore, one should use considerable caution when deriving conclusions based on \mbh\ estimates derived from this method, which should not be used in isolation.

\end{itemize}

\section*{Acknowledgements}
We thank the referee for the constructive comments and suggestions that have improved the clarity of this paper.

\section*{Data Availability}
The data underlying this article are available in the High Energy Astrophysics Science Archive Research Center (HEASARC) archive at https://heasarc.gsfc.nasa.gov/docs/archive.html.



\bibliographystyle{mnras}
\bibliography{MBHindirect} 




\appendix

\section{NuSTAR observations and spectral results}
Table~\ref{tab:obsids} shows the \nustar\ observations used in this paper. Table~\ref{tab:spectral} shows the spectral results obtained by the \textsc{xspec} modeling described in Section~\ref{sec:X-ray spectral}.

\begin{table*}
	\caption{\nustar\ observations used in this paper}	
	\begin{center}
	\begin{tabular}{lrrrl} 
			\toprule
			\toprule
			\mcol{Name} & \mcol{ObsID} & 
            \mcol{Date} & \mcol{Exposure} & \mcol{Analysis} \\
			  & & & \mcol{(s)} \\
    		\midrule
			NGC 678 & 60760001002 & 2020-12-25 & 29513 & spec \\
			NGC 1052 & 60201056002 & 2017-01-17 & 59754 & spec \\
			NGC 1068 & 60002030002 & 2012-12-18 & 57850 & var \\
			& 60002030004 & 2012-12-20 & 48556 & var \\
			& 60002033002 & 2014-08-18 & 52055 & spec \& var \\
			& 60002033004 & 2015-02-05 & 53685 & var \\
			& 60302003002 & 2017-07-31 & 49979 & var \\
			& 60302003004 & 2017-08-27 & 52549 & var \\
			& 60302003006 & 2017-11-06 & 49691 & var \\
			& 60302003008 & 2018-02-05 & 54624 & var \\
			NGC 1365 & 60002046007 & 2013-01-23 & 73648 & var \\
			& 60002046009 & 2013-02-12 & 69871 & var \\
			& 60702058002 & 2021-04-16 & 56297 & spec \\
			NGC 1566 & 60501031002 & 2019-08-08 & 58922 & spec \& var \\
			& 60501031004 & 2019-08-18 & 77200 & var \\
			& 60501031006 & 2019-08-21 & 86047 & var \\
			& 80301601002 & 2018-06-26 & 56836 & var \\
			& 80401601002 & 2018-10-04 & 75395 & var \\
			& 80502606002 & 2019-06-05 & 57262 & var \\
			NGC 2110 & 60061061002 & 2012-10-05 & 15535 & spec \\
			ESO 5-4 & 60662006002 & 2021-05-29 & 22054 & spec \\
			NGC 2992 & 90501623002 & 2019-05-10 & 57490 & spec \\
			NGC 3079 & 60061097002 & 2013-11-12 & 21541 & spec \\
			NGC 3081 & 60561044002 & 2019-12-23 & 55634 & spec \& var \\
			NGC 3227 & 60202002002 & 2016-11-09 & 49800 & var \\
			& 60202002004 & 2016-11-25 & 42457 & spec \& var \\
			& 60202002008 & 2016-12-01 & 41812 & var \\
			& 60202002010 & 2016-12-05 & 40887 & var \\
			& 60202002014 & 2017-01-21 & 47602 & var \\
			NGC 3516 & 60002042002 & 2014-06-24 & 50998 & var \\
			& 60002042004 & 2014-07-11 & 72088 & var \\
			& 60160001002 & 2020-04-20 & 39857 & spec \\
			NGC 3783 & 60101110002 & 2016-08-22 & 41265 & var \\
			& 60101110004 & 2016-08-24 & 42428 & spec \& var \\
			NGC 4151 & 60001111002 & 2012-11-12 & 21861 & spec \\
			& 60001111005 & 2012-11-14 & 61528 & var \\
			& 60502017004 & 2019-11-12 & 43738 & spec \& var \\
			& 60502017006 & 2019-12-24 & 32401 & spec \\
			& 60502017012 & 2020-01-23 & 28859 & spec \\
			NGC 4388 & 60061228002 & 2013-12-27 & 21384 & spec \\
			& 60501018002 & 2019-12-25 & 50360 & var \\
			NGC 4500 & 60510002002 & 2019-07-03 & 55168 & spec \\
			NGC 4593 & 60001149002 & 2014-12-29 & 23317 & spec \\
			NGC 4945 & 60002051004 & 2013-06-15 & 54614 & spec \\
			NGC 5290 & 60160554002 & 2021-07-28 & 18812 & spec \\
			Circinus & 60002039002 & 2013-01-25 & 53873 & spec \\
			NGC 5506 & 60501015002 & 2019-12-28 & 61384 & spec \& var \\
			& 60501015004 & 2020-02-09 & 47480 & var \\
			NGC 5728 & 60061256002 & 2013-01-02 & 24357 & spec \\
			ESO 137-34 & 60061272004 & 2020-04-12 & 28128 & spec \\
			NGC 6221 & 60160651002 & 2016-05-23 & 18470 & spec \\
            \multicolumn{5}{c}{\textit{continued}} \\
			\bottomrule
		\end{tabular}	
	\end{center}
	\label{tab:obsids}
\end{table*}

\begin{table*}
\ContinuedFloat
	\caption{\textit{(continued)}}	
	\begin{center}
	\begin{tabular}{lrrrl} 
			\toprule
			\toprule
			\mcol{Name} & \mcol{ObsID} & 
            \mcol{Date} & \mcol{Exposure} & \mcol{Analysis} \\
			  & & & \mcol{(s)} \\
    		\midrule
			NGC 6300 & 60261001004 & 2016-08-24 & 23541 & spec \\
			NGC 6814 & 60201028002 & 2016-07-04 & 148428 & spec \\
			& 60701012002 & 2021-10-01 & 128219 & var \\
			NGC 7172 & 60061308002 & 2014-10-07 & 32001 & spec \\
			NGC 7213 & 60001031002 & 2014-10-05 & 101612 & spec \& var \\
			NGC 7314 & 60201031002 & 2016-05-13 & 100422 & spec \& var \\
			NGC 7465 & 60160815002 & 2020-01-09 & 20983 & spec \\
			  NGC 7479 & 60061316002 & 2020-11-06 & 23646 & spec \\
            NGC 7582 & 60201003002 & 2016-04-28 & 48488 & spec \\
			\bottomrule
		\end{tabular}	
	\end{center}
	\begin{flushleft}
        In the Analysis column, \textit{spec} denotes objects we conducted spectral analysis on, and for those we calculated masses with the X-ray scaling method. In the same column, \textit{var} denotes objects we conducted temporal analysis on, and for those we calculated masses with the X-ray variability method.
	\end{flushleft}
	\label{tab:obsids-cont}
\end{table*}

A detailed description of the spectral analysis and results of NGC 1068, NGC 3079, NGC 4388, NGC 4945, and Circinus is provided in \citet{Gliozzi2021}. Similarly, the detailed spectral analysis of NGC 3081, NGC 4593, NGC 6814, NGC 7172, NGC 7314, and NGC 7465 is reported in \citet{Shuvo2022}, and that of NGC 4151 is described in \citet{Williams2023}. All the other sources were analyzed and systematically fitted with the baseline model described in Sec.~\ref{sec:X-ray spectral}, which yielded reasonably good fits for the vast majority of them. The few sources that required some additional spectral components are reported below. 

\noindent NGC 3516 includes a Gaussian component \texttt{zgauss} at E=6 keV.

\noindent NGC 5728 includes an additive component \texttt{const*BMC} to parametrize the fraction of primary X-ray emission directly scattered towards the observer.

\noindent NGC 7582 also includes an additive component \texttt{const*BMC}.

\begin{table*}
	\caption{Spectral results}		
	\begin{center}
	\begin{tabular}{lrrrrrrrrr} 
			\toprule
			\toprule
			\mcol{Name} & \mcol{log($N_{\textrm{H},\textrm{bor}}$)} & 
            \mcol{CF$_\textrm{bor}$} & \mcol{$N_{\textrm{H},\textrm{los}}$} & 
            \mcol{$\Gamma$} & \mcol{\nbmc} & 
            \mcol{k$T$} & \mcol{log $A$} & 
            \mcol{$\chi^2_\textrm{red}$/d.o.f.} & \mcol{$L_\textrm{2--10 keV}$} \\
			  & & & 
            \mcol{(cm$^{-2}$)} & & 
            & \mcol{(keV)} & 
            & & \mcol{(erg s$^{-1}$)} \\
			  \mcol{(1)} & \mcol{(2)} & \mcol{(3)} & \mcol{(4)} & 
            \mcol{(5)} & \mcol{(6)}& \mcol{(7)} & \mcol{(8)} & 
            \mcol{(9)} & \mcol{(10)} \\
			\midrule
			NGC 678 & $24.2\pm0.1$ & 0.89 & $(3.7\pm0.2) \times 10^{23}$ & 
            $1.93^{+0.06}_{-0.08}$ & $3.5^{+0.8}_{-0.8} \times 10^{-5}$ & 
            0.05 & 0.63 & 
            0.86/91 & $5.36 \times 10^{41}$ \\
            \noalign{\smallskip}
			NGC 1052 & $23.5\pm0.1$ & 0.73 & $(1.1\pm0.1) \times 10^{23}$ & 
            $1.68^{+0.01}_{-0.01}$ & $8.6^{+0.2}_{-0.2} \times 10^{-5}$ & 
            0.14 & 0.44 & 
            1.00/729 & $4.61 \times 10^{41}$ \\
            \noalign{\smallskip}
			NGC 1068$^\clubsuit$ & $23.3\pm0.1$ & 0.15 & $(3.5\pm0.1) \times 10^{24}$ & 
            $1.98^{+0.01}_{-0.01}$ & $2.2^{+0.1}_{-0.1} \times 10^{-3}$ & 
            0.1 & 0.08 & 
            1.06/713 & $3.31 \times 10^{42}$ \\
            \noalign{\smallskip}
			NGC 1365 & $24.4\pm0.1$ & 0.91 & 
            $(2.2\pm0.1) \times 10^{23}$ & $1.69^{+0.01}_{-0.01}$ &
            $1.4^{+0.1}_{-0.1} \times 10^{-4}$ & 0.20 &
            0.14 & 0.97/828 & $5.41 \times 10^{41}$ \\
            \noalign{\smallskip}
			NGC 1566 & $24.6 \pm 0.1$ & 0.49 & 
            $(8.1\pm8.1) \times 10^{20}$ & $1.82^{+0.02}_{-0.02}$ &
            $1.1^{+0.1}_{-0.1} \times 10^{-4}$ & 0.01 &
            0.51 & 0.98/916 & $7.16 \times 10^{41}$ \\
            \noalign{\smallskip}
			NGC 2110 & $22.9 \pm 0.3$ & $0.15$ & 
            $(4.3 \pm 0.2) \times 10^{22}$ & $1.71^{+0.01}_{-0.01}$ &
            $3.0^{+0.1}_{-0.1} \times 10^{-3}$ & 0.25 &
            0.33 & 0.96/1216 & $3.89 \times 10^{43}$ \\
            \noalign{\smallskip}
			ESO 5-4 & $23.9 \pm 0.1$ & $0.91$ & 
            $(4.3 \pm 0.6) \times 10^{23}$ & $1.45^{+0.06}$ &
            $3.7^{+0.6}_{-0.1} \times 10^{-5}$ & 0.72 &
            -7.18 & 1.26/62 & $1.94 \times 10^{41}$ \\
            \noalign{\smallskip}
			NGC 2992 & $23.5 \pm 0.1$ & $0.91$ & 
            $(1.2 \pm 0.1) \times 10^{22}$ & $1.77^{+0.01}_{-0.01}$ &
            $5.8^{+0.1}_{-0.1} \times 10^{-4}$ & 0.03 &
            0.40 & 0.98/1505 & $1.51 \times 10^{43}$ \\
            \noalign{\smallskip}
			NGC 3081$^\diamondsuit$ & $23.1\pm0.1$ & 0.90 & $(0.8\pm0.1) \times 10^{24}$ & 
            $1.72^{+0.01}_{-0.01}$ & $4.9^{+0.2}_{-0.2} \times 10^{-4}$ & 
            0.10 & 0.50 & 
            1.01/863 & $8.25 \times 10^{42}$ \\
            \noalign{\smallskip}
			NGC 3079$^\clubsuit$  & $24.5\pm0.1$ & 0.20 & $(2.9\pm0.1) \times 10^{24}$ & 
            $1.91^{+0.05}_{-0.06}$ & $1.1^{+0.2}_{-0.2} \times 10^{-3}$ & 
            0.10 & 0.80 & 
            1.07/75 & $8.36 \times 10^{42}$ \\
            \noalign{\smallskip}
			NGC 3227 & $24.2\pm0.1$ & 0.60 & 
            $(1.6\pm0.3)\times10^{22}$ & $1.80^{+0.01}_{-0.01}$ &
            $3.4^{+0.1}_{-0.1}\times10^{-4}$ & 0.35 &
            0.50 & 0.99/1017 & $1.97\times10^{42}$ \\
            \noalign{\smallskip}
			NGC 3516 & $24.2\pm0.1$ & 0.91 & 
            $(2.5\pm0.1)\times10^{22}$ & $1.93^{+0.01}_{-0.01}$ &
            $4.7^{+0.1}_{-0.1}\times10^{-4}$ & 0.36 &
            0.65 & 1.01/1045 & $8.04\times10^{42}$ \\
            \noalign{\smallskip}
			NGC 3783 & $24.2\pm0.1$ & 0.91 & 
            $(4.0\pm0.1)\times10^{22}$ & $1.89^{+0.01}_{-0.01}$ &
            $5.5^{+0.1}_{-0.1}\times10^{-4}$ & 0.36 &
            0.23 & 1.07/1129 & $7.62\times10^{42}$ \\
            \noalign{\smallskip}
			NGC 4151 & $24.0\pm0.1$ & 0.91 & $(18.6\pm0.4) \times 10^{24}$ & 
            $1.83^{+0.01}_{-0.01}$ & $2.2^{+0.1}_{-0.1} \times 10^{-3}$ & 
            0.11 & 1.45 & 
            1.07/1563 & $8.3 \times 10^{42}$ \\
            \noalign{\smallskip}
			NGC 4388$^\clubsuit$ & $23.6\pm0.1$ & 0.91 & $(4.0\pm1.0) \times 10^{23}$ & 
            $1.66^{+0.04}_{-0.04}$ & $3.3^{+0.5}_{-0.4} \times 10^{-4}$ & 
            0.10 & -0.55 & 
            1.04/420 & $5.41 \times 10^{41}$ \\
            \noalign{\smallskip}
			NGC 4500 & $24.0 \pm 0.1$ & $0.84$ & 
            $(3.0 \pm 0.1) \times 10^{23}$ & $1.50^{+0.03}_{-0.03}$ &
            $1.8^{+0.1}_{-0.1} \times 10^{-5}$ & 0.17 &
            0.26 & 0.93/177 & $2.92 \times 10^{41}$ \\
            \noalign{\smallskip}
			NGC 4593$^\diamondsuit$ & $25.5\pm0.5$ & 0.36 & $(2.5\pm1.5) \times 10^{20}$ & 
            $1.88^{+0.01}_{-0.01}$ & $2.3^{+0.1}_{-0.1} \times 10^{-4}$ & 
            0.10 & 0.20 & 
            1.01/678 & $3.82 \times 10^{42}$ \\
            \noalign{\smallskip}
			NGC 4945$^\clubsuit$ & $24.4\pm0.1$ & 0.91 & $(3.0\pm0.7) \times 10^{24}$ & 
            $1.74^{+0.05}_{-0.05}$ & $1.6^{+0.1}_{-0.1} \times 10^{-3}$ & 
            0.10 & 2.15 & 
            0.99/1716 & $5.55 \times 10^{41}$ \\
            \noalign{\smallskip}
			NGC 5290 & $24.8 \pm 0.1$ & $0.58$ & 
            $(3.7 \pm 0.5) \times 10^{22}$ & $1.86^{+0.03}_{-0.03}$ &
            $9.8^{+0.5}_{-0.5} \times 10^{-5}$ & 0.19 &
            0.06 & 1.00/185 & $9.39 \times 10^{41}$ \\
            \noalign{\smallskip}
			Circinus$^\clubsuit$ & $23.6\pm0.1$ & 0.24 & $(1.6\pm0.1) \times 10^{24}$ & 
            $2.17^{+0.01}_{-0.01}$ & $9.8^{+0.1}_{-0.2} \times 10^{-3}$ & 
            0.10 & -0.43 & 
            1.00/1714 & $4.50 \times 10^{41}$ \\
            \noalign{\smallskip}
			NGC 5506 & $24.0\pm0.1$ & 0.91 & 
            $(4.7\pm0.1)\times10^{22}$ & $1.86^{+0.01}_{-0.01}$ &
            $8.0^{+0.1}_{-0.1}\times10^{-4}$ & 0.47 &
            0.40 & 0.97/1410 & $5.72\times10^{42}$ \\
            \noalign{\smallskip}
			NGC 5728 & $24.2\pm0.1$ & 0.90 & 
            $(8.9\pm0.2)\times 10^{23}$ & $1.72^{+0.02}_{-0.02}$ &
            $2.5^{+0.1}_{-0.1} \times 10^{-4}$ & 0.14 &
            0.32 & 0.97/365 & $4.42 \times 10^{42}$ \\
            \noalign{\smallskip}
			ESO 137-34 & $24.2\pm0.1$ & 0.91 & 
            $(1.6\pm0.2)\times10^{23}$ & $1.59^{+0.02}_{-0.01}$ &
            $1.3^{+0.1}_{-0.1}\times10^{-4}$ & 0.34 &
            -1.26 & 1.12/195 & $3.49\times10^{41}$ \\
            \noalign{\smallskip}
			NGC 6221 & $23.8\pm0.1$ & 0.91 & 
            $(1.0\pm0.8)\times10^{22}$ & $1.85^{+0.03}_{-0.03}$ &
            $7.0^{+0.5}_{-0.5}\times10^{-5}$ & 0.10 &
            0.19 & 0.97/232 & $1.02\times10^{41}$ \\
            \noalign{\smallskip}
			NGC 6300 & $24.2\pm0.1$ & 0.91 & 
            $(1.6\pm0.2)\times10^{23}$ & $1.91^{+0.01}_{-0.01}$ &
            $4.2^{+0.1}_{-0.1}\times10^{-4}$ & 0.10 &
            0.54 & 0.93/793 & $8.69\times10^{41}$ \\
            \noalign{\smallskip}
			NGC 6814$^\diamondsuit$ & $24.3\pm0.1$ & 0.46 & $(8.0\pm1.0) \times 10^{21}$ & 
            $1.88^{+0.01}_{-0.01}$ & $3.0^{+0.1}_{-0.1} \times 10^{-4}$ & 
            0.10 & 0.5 & 
            1.04/1520 & $2.17 \times 10^{42}$  \\
            \noalign{\smallskip}
			NGC 7172$^\diamondsuit$ & $24.1\pm0.1$ & 0.80 & $(1.0\pm0.1) \times 10^{23}$ & 
            $1.90^{+0.02}_{-0.01}$ & $6.7^{+0.9}_{-0.6} \times 10^{-4}$ & 
            0.10 & 1.7 & 
            1.00/1150 & $1.16 \times 10^{43}$ \\
            \noalign{\smallskip}
			NGC 7213 & $23.4 \pm 0.1$ & 0.80 & 
            $(7.8 \pm 1.1) \times 10^{21}$ & $1.93^{+0.01}_{-0.01}$ &
            $1.7^{+0.1}_{-0.1} \times 10^{-4}$ & 0.05 &
            0.58 & 1.00/921 & $9.56 \times 10^{41}$ \\
            \noalign{\smallskip}
			NGC 7314$^\diamondsuit$ & $24.2\pm0.1$ & 0.79 & $(1.2\pm0.1) \times 10^{22}$ & 
            $2.03^{+0.01}_{-0.01}$ & $4.9^{+0.1}_{-0.1} \times 10^{-4}$ & 
            0.10 & 0.5 & 
            1.03/1276 & $1.16 \times 10^{43}$  \\
            \noalign{\smallskip}
			NGC 7465$^\diamondsuit$ & $23.8\pm0.1$ & 0.91 & $(1.0\pm0.5) \times 10^{22}$ & 
            $1.87^{+0.02}_{-0.02}$ & $1.2^{+0.1}_{-0.1} \times 10^{-4}$ & 
            0.10 & 0.5 & 
            0.96/465 & $1.13 \times 10^{42}$ \\
            \noalign{\smallskip}
			NGC 7479 & $24.4\pm0.1$ & 0.91 & 
            $(4.9\pm0.4)\times10^{23}$ & $1.62^{+0.05}_{-0.03}$ &
            $9.3^{+1.1}_{-0.8}\times10^{-5}$ & 0.61 &
            -1.28 & 0.97/70 & $7.21\times10^{41}$ \\
            \noalign{\smallskip}
			NGC 7582 & $24.2\pm0.1$ & 0.70 & 
            $(3.2\pm0.1)\times10^{23}$ & $1.78^{+0.01}_{-0.01}$ &
            $4.0^{+0.1}_{-0.1}\times10^{-4}$ & 0.29 &
            0.54 & 1.06/1028 & $2.39\times10^{42}$ \\
			\bottomrule
		\end{tabular}	
	\end{center}
	\begin{flushleft}
		Columns. 1 = object name (the club suit indicates objects whose X-ray spectral analysis was performed in \citet{Gliozzi2021}, the diamond suit refers to objects analyzed in \citet{Shuvo2022}, and the analysis of NGC 4151 was carried out in \citet{Williams2023}). 2 = log of the column density computed by the \textsc{borus} model. 3 = covering factor computed by \textsc{borus}. 4 = line-of-sight column density computed by \textsc{MYTZ} or \textsc{zphabs}. 5 = photon index. 6 = \textsc{bmc} normalization, defined as the accretion luminosity in units of $10^{39}$ erg s$^{-1}$ divided by the distance squared in units of 10 kpc. 7 = temperature of the thermal seed photons in keV. 8 = log of the $A$ parameter, which is related to the Comptonized fraction of photons $f$ by the equation $f=A/(A+1)$. 9 = reduced $\chi^2$ and degrees of freedom of the fit. 10 = X-ray luminosity in the 2--10 keV band.
	\end{flushleft}
	\label{tab:spectral}
\end{table*}

\bsp	
\label{lastpage}
\end{document}